\documentclass[lettersize,journal]{IEEEtran}
\usepackage{amsmath,amsfonts}
\usepackage{algorithmic}
\usepackage{algorithm}
\usepackage{array}
\usepackage[caption=true,font=normalsize,labelfont=sf,textfont=sf]{subfig}
\usepackage{textcomp}
\usepackage{stfloats}
\usepackage{url}
\usepackage{verbatim}
\usepackage{graphicx}
\usepackage{cite}
\usepackage{diagbox}
\usepackage{slashbox}

\usepackage{booktabs}
\usepackage{caption}
\usepackage{filecontents}
\usepackage{float}
\usepackage{adjustbox}
\usepackage{multirow}
\usepackage{url}
\usepackage{xcolor}

\begin{document}

\title{Classification and Explanation of Distributed Denial-of-Service (DDoS) Attack Detection using Machine Learning and Shapley Additive Explanation (SHAP) Methods}
% Applying Machine Learning-based Classification and Explanation with shapley additive explanation(SHAP) for DDoS Detection
\author{Yuanyuan Wei, Julian Jang-Jaccard, Amardeep Singh, Fariza Sabrina, ~\IEEEmembership{Member,~IEEE}, and Seyit Camtepe, ~\IEEEmembership{Senior Member,~IEEE}
        % <-this % stops a space
\thanks{Yuanyuan Wei is with the CybersecurityLab, Comp Sci/Info Tech, Massey University, Auckland, 0632, NEW ZEALAND (e-mail: y.wei1@massey.ac.nz).}
\thanks{Julian Jang-Jaccard is with the CybersecurityLab, Comp Sci/Info Tech, Massey University, Auckland, 0632, NEW ZEALAND (e-mail: j.jang-jaccard@massey.ac.nz).}
\thanks{Amardeep Singh is with the UCOL Te Pūkenga, New Zealand (e-mail: a.singh@ucol.ac.nz) }
\thanks{Fariza Sabrina is with the School of Engineering and Technology, Central Queensland University, Sydney NSW 2000, AUSTRALIA (e-mail: f.sabrina@cqu.edu.au).}
% \thanks{Wen Xu is with the CybersecurityLab, Comp Sci/Info Tech, Massey University, Auckland, 0632, NEW ZEALAND (e-mail: w.xu2@massey.ac.nz).}
\thanks{Seyit Camtepe is with the CSIRO Data61, AUSTRALIA (e-mail: Seyit.Camtepe@data61.csiro.au).}

\thanks{Manuscript received April 19, 2021; revised August 16, 2021.}}

\markboth{Journal of \LaTeX\ Class Files,~Vol.~14, No.~8, August~2021}%
{Shell \MakeLowercase{\textit{et al.}}: A Sample Article Using IEEEtran.cls for IEEE Journals}


\maketitle

\begin{abstract}
DDoS attacks involve overwhelming a target system with a large number of requests or traffic from multiple sources, disrupting the normal traffic of a targeted server, service, or network. Distinguishing between legitimate traffic and malicious traffic is a challenging task. It is possible to classify legitimate traffic and malicious traffic and analysis the network traffic by using machine learning and deep learning techniques. Moreover, machine learning techniques are faster to identify DDoS attacks and allow for efficient response and mitigation. However, an inter-model explanation implemented to classify a traffic flow whether is benign or malicious is an important investigation of the inner working theory of the model to increase the trustworthiness of the model. Explainable Artificial Intelligence (XAI) can explain the decision-making of the machine learning models that can be classified and identify DDoS traffic. In this context, we proposed a framework that can not only classify legitimate traffic and malicious traffic of DDoS attacks but also use SHAP to explain the decision-making of the classifier model. To address this concern, we first adopt feature selection techniques to select the top 20 important features based on feature importance techniques (e.g., XGB-based feature importance, Permutation feature importance, and SHAP feature importance). Following that, the Multi-layer Perceptron Network (MLP) part of our proposed model uses the optimized features of the DDoS attack dataset as inputs to classify legitimate traffic and malicious traffic. We perform extensive experiments with all features and selected features in one-to-one (benign vs. one attack type) and one-to-all (benign vs. multi-attack types) scenarios. The evaluation results show that the model performance with selected features achieves above 99\% accuracy in both one-to-one and one-to-all classifications. Finally, to provide interpretability, XAI can be adopted to explain the model performance between the prediction results and features based on global and local explanations by SHAP, which can better explain the results achieved by our proposed framework.

\end{abstract}


\section{Introduction}
\IEEEPARstart{C}{y}bersecurity refers to the aspect of protecting the individual or organization's network and information systems to defend against cyber attacks, including protecting data availability, confidentiality, and integrity~\cite{vsarvcevic2022cybersecurity}. In the era of rapid information technology development, network security is becoming increasingly important to networks and computer users.  However, the rapid proliferation of innovative technologies and communication infrastructure brings the potential for cyberattacks and other threats to Internet users~\cite{haider2020deep,sahoo2019toward,yuan2017deepdefense}. One of the significant cyber attacks is the distributed denial-of-service (DDoS) attack. A DDoS attack is a cyberattack that poses serious risks to the network system service. A DDoS attack is a kind of malicious traffic attempt to disrupt the normal traffic of the targeted server or network service by sending a flood of malicious traffic, making the target services unavailable to legitimate users\cite{koay2018new,doriguzzi2020lucid,salahuddin2020time,aydin2022long}. Cyber-attacks, especially DDoS attacks can be used to maliciously disable computers, prevent access to them, steal data, or use a compromised computer as a launching pad for other attacks\cite{aydin2022long}. As a result, preventing and mitigating DDoS attacks  is critical to protecting the network  from significant financial loss~\cite{salahuddin2021chronos,elsayed2020ddosnet}.

%the era of rapid development of information technology, network security is becoming increasingly important with the development of various applications such as 5G communication technology, big data, cloud computing, and the Internet of Things (IoT), etc. An estimated one trillion physical devices were connected to the internet last year (2022)~\cite{santos2018intrusion}. In addition, it is estimated that 29 billion IoT devices will be connected and deployed globally by 2030, with significant use across all types of industry verticals and consumer markets, including smartphones, autonomous vehicles, IT infrastructure, asset tracking and monitoring, and smart grid~\cite{Lionel2022number}. 
% However, as new technologies continue to develop rapidly, many cybersecurity and privacy concerns are being raised. For instance, widely used and open Internet devices are easy targets for cyber-attacks. 

In recent years, machine learning (ML) has been proposed and developed by many researchers for the detection of DDoS attacks. For example, K-Nearest Neghbors (KNN) \cite{dong2019ddos}, Random Forest (RF) \cite{chen2020ddos}, Extreme Gradient Boosting (XGBoost)~\cite{chen2018xgboost}, and Decision Tree (DT)~\cite{kousar2021detection,khare2020real}. In addition, Deep Learning (DL) techniques have also been used to detect DDoS attacks. These include long short-term memory (LSTM)~\cite{aydin2022long}, Autoencoder (AE)~\cite{salahuddin2021chronos}, AE-MLP~\cite{wei2021ae}, Convolutional Neural Network (CNN)~\cite{haider2020deep}, etc. As aforementioned, many researchers achieve power in terms of performance and prediction by using AI techniques, especially ML learning and DL learning in terms of application in the cybersecurity domain, for example, DDoS attack detection, intrusion detection, and malicious detection based on benchmark datasets such as NSL-KDD, CICIDS 2017, CICDDoS2019 datasets~\cite{idhammad2018semi,rajagopal2021towards,akgun2022new,khoei2021ensemble,novaes2021adversarial}. This AI-based ML and DL techniques for DDoS detection can be divided into two categories: white box and black box models. The white-box decision-making of some of the ML models can be easily explained and trusted by domain experts based on the interpretable output like linear models.~\cite{lundberg2017unified,sharma2022explainable}. However, despite the good performance of the AI-based model, the domain experts were still unable to understand the inner workings of some AI-based models and fully trust them, resulting in a potential loss of security and trust. These models can be referred to as black-box models in ML and DL models. With the complexity of the AI models being developed, understanding the good performance of the decisions made by AI-based models has become an important issue so that domain experts can gain more confidence in the black-box models. Therefore, it is important to make a correct and transparent interpretative prediction. To address this issue, Explainable Artificial Intelligence (XAI) is proposed to investigate the interpretability of the inner logic of the model and to explain the decision-making behind the model, thereby making the model transparent to user understanding.

% In the case of DDoS detection, the emergence of a new type of attack has been accompanied by a massive increase in Internet traffic data. For DDoS detection, such as classification or anomaly detection, the AI techniques may not correctly identify these new attack types (e.g. zero attack), resulting in the attacks being misclassified as normal traffic. XAI is an essential technology that can provide cybersecurity with a transparent explanation of the decision made by ML or DL models in order to detect more unknown attacks.

In this paper, we propose a framework using the Shapley Additive Explanations (SHAP) technique to address cybersecurity issues such as misclassification and provide a transparent explanation for DDoS detection. The goal of this paper is twofold. First, SHAP feature importance is used as a feature extraction to select the top N contributing features that are fed into the classifier model in order to obtain the accuracy and misclassifications of the predictions. Second, the XAI-based technique, such as SHAP, provides global and local interpretations that can explain the contribution of the features for the decision-making of the model. 
% transparency of DDoS.

The main contributions of our proposed model are the following. 

\begin{itemize}
    \item We build a combination feature selection framework based on three XGB-based feature importance techniques that select the top n relevant importance features, which can not only reduce the computational time but also improve the model efficiency.
    \item We propose a novel framework that consists of two components: DDoS classification using MLP and XAI-based technique of SHAP for transparency to explain the most contributing features of the classification of the benign and attack traffics. This work helps the users trust and have a better understanding of the prediction results of the proposed model.
    \item We propose the XAI-based SHAP interpretation framework for the global and local explanation of the classification result in DDoS attack detection. The global explanation we conduct is based on summary plots and dependence plots. For the local explanation, we focus the analysis on a single traffic sample in four circumstances, which are benign and malicious correctly traffic classified, and misclassified traffic, while other studies only provide the analysis on misclassified traffic.
    \item We conducted an extensive evaluation using the CICDDoS2019 dataset to find the most corresponding features, with the explanation attached as the best exploration of the most contributing features. The evaluation results show that using the feature selected by the XAI feature importance technique, the classification results show higher performance than using all features.

	% \item We first use three feature importance schemes to select the top 10 important features, and then feed  them into the classifier model to obtain the prediction results.
	% % \item We compare the results based on MLP (or XGBoost) classifiers by exploring the differences between different feature extraction techniques.
	% \item To explore the decision behind the trained mode, we provide XAI-based techniques of SHAP to explain the proposed framework based on global and local interpretation. Therefore, the proposed framework can increase human confidence in DDoS detection.
	% \item The state-of-the-art benchmark dataset of the CICDDoS2019 dataset can be used in this implementation and comparison. 
\end{itemize}

The rest of this paper is structured as follows: Section~\ref{sec:rw} introduces related works in the field of DDoS attack detection.  Section~\ref{sec:method} introduces our methodology. Section~\ref{sec:re} illustrates the experimental setup and  Section \ref{sec:Experiments} details the analysis of our results and evaluated the global and local explanations. Section~\ref{sec:conclusion} concludes the paper with the planned future works.

\section{Related worked} \label{sec:rw}
In recent years, ML and DL techniques have been widely used to detect DDoS attacks, analyze network traffic patterns, and detect normal or abnormal behavior that may detect DDoS traffic efficiently. However, these techniques are commonly referred to as black-box models, which can make it difficult to understand the inner workings of the decisions behind them. XAI is an area of research that aims to make the results of AI-driven models more transparent and easier for domain experts to understand. This section discusses some existing work that addresses DDoS attack detection and explainability.

\subsection{Machine learning for DDoS}
Wang et al.~\cite{wang2020dynamic} used a dynamic multilayer perceptron (MLP) with 31 optimized sequence features and feedback mechanism to detect DDoS attacks based on the NSL-KDD dataset, achieving 97.66\% accuracy with the SBS-MLP classifier. Wei et al.~\cite{wei2021ae} proposed a hybrid deep learning Autoencoder-MLP (AE-MLP) for DDoS detection and classification. They first used the autoencoder to extract 5 optimal features, which were fed into MLP classifiers to perform multi-class classification of different attack types on the CICDDoS2019 dataset. The evaluation results achieved over 98\% accuracy in classifying all attack types. In addition, Can et al.~\cite{can2021detection} proposed Distributed Denial of Service attacks based on neural networks (DDoSNet) to detect and classify DDoS attacks based on a fully-connected MLP classifier with 24 selected features from the CICDDoS2019 dataset. The proposed method achieved 99\% accuracy for binary classification. Samom et al.~\cite{singh2021distributed} used machine learning models (i.e., Logistic Regression, Random Forest, Multi-Layer Perceptron, etc) to classify four different attack types (i.e., SYN, NET, Portmap, and UDPLag) with 20 selected features from CICDDoS2019 dataset. The results showed that Random Forest demonstrated the best performance in classification results. The aforementioned machine learning techniques achieved higher performance accuracy for DDoS detection and classification with selected features, but lower performance as they used the entire feature set for classification. 

\subsection{XAI for models}
% Batchu et al.~\cite{batchu2022integrated} has developed an efficient model to improve classification decision-making, achieving high accuracy (99\%), and making the results more transparent through SHAP and LIME explainer for the global and local explanation. They used the Adaptive Synthetic oversampling technique to solve the problem of unbalanced data and applied SHAP feature importance to extract the best features for better classification detection. However, this research only worked in binary classification. 

Kalutharage et al.~\cite{kalutharage2023explainable} proposed XAI-based techniques for detecting DDoS attack anomalies on the USBIDS (University of Sannio, Benevento Instrution Detection System) dataset. The research focuses on instance-by-instance, local and global explanations, and feature correlations, explaining anomalies by providing Autoencoder and Kernel SHAP techniques. The proposed method achieves a better accuracy (98\%) on HULK attacks (Hulk No Defense) compared to other detection methods, such as Decision Tree (97\%), and Deep Neural Network (67\%). However, this research was implemented in a static dataset. Antwarg et al.~\cite{antwarg2021explaining} proposed Kenal SHAP for explaining anomalies  on the NSL-KDD dataset by an unsupervised model of Autoencodr. The main focus was on explaining to the experts the relationship between the impact of reconstruction error features and high reconstruction error features. The explanation presented the main contribution and offset the important features of the anomalies. The evaluation of the explanation also showed that utilizing the SHAP explanation technique to explain the generated subset of explanatory features is more robust than the other explanation techniques of LIME.  Šarčević et al.~\cite{vsarvcevic2022cybersecurity} provided both SHAP's XAI technique and If-then decision tree rules to extract information from a network attack dataset CIC-IDS2017 to mitigate network security. The challenges of these two techniques for information extraction were compared and their use in different situations was pointed out. The limitations of both techniques were also highlighted. If-then decision tree rules face the challenge of increasing the depth of the trees and cannot provide transparency in the decision-making of the results, while SHAP provides an adequate explanation of the model but is less comprehensive.

Tabassun et al.~\cite{tabassum2022iot} extended their work by employing XAI techniques of SHAP, LIME, and ELI5 to explain the classification of IoT network attacks (DDoS attacks) based on machine learning and deep learning models (e.g., DT, RF, Adaboost, ANN, etc). XAI explanation techniques are based on the decision-making of the binary classification results of 96\% accuracy of all models. Using SHAP for a both local and global explanation of the prediction results based on SHAP values,  while LIME provided only a local explanation, and ELI5 pointed out the most important features both locally and globally. Houda et al.~\cite{abou2022should} proposed an XAI framework to explain the decision of deep learning for Internet of Things (IoT)-related Intrusion Detection Systems (IDSs). They also provide three XAI techniques, including SHAP, LIME, and RuleFit, to optimize the interpretability of deep learning decisions through global and local explanations.
% Sharma et al.~\cite{sharma2022explainable} presents a better solution for a more accurate explanation of the decision-making process of misclassification. They use 4 classifiers (random forest, K-Nearest Neighbors, Multi-Layer Perceptron, and Support Vector Machine) to evaluate the results.

In this study, we propose a framework for classifying the legitimate traffic (benign) and malicious traffic (attack) based on a machine learning technique of MLP classifier and give an interpretable to the trained model using XAI-based technique of SHAP based on selected top 20 features. Compared with The aforementioned \cite{wang2020dynamic,wei2021ae,can2021detection}, which used a different feature selection technique to select the top N features, we proposed a feature selection framework by selecting the top 20 important features based on the combination of XGB-based feature importance, Permutation importance SHAP feature importance with each feature occurrence frequency.

%%%%%%%%%%%%%%%%%%
% \textbf{Three papers}
% \begin{itemize}
%     \item An integrated approach explaining the detection of distributed denial of service attacks \cite{batchu2022integrated}
%     \item Explaining anomalies detected by autoencoders using Shapley Additive Explanations \cite{antwarg2021explaining}
%     \item An explainable machine learning framework for intrusion detection systems \cite{wang2020explainable}
% \end{itemize}

% \textbf{Survey Paper}
% \begin{itemize}
%     \item Peeking Inside the Black-Box: A Survey on Explainable Artificial Intelligence (XAI) \cite{adadi2018peeking}
%     \item Explainable Artificial Intelligence (XAI): Concepts, taxonomies, opportunities and challenges toward responsible AI \cite{arrieta2020explainable}
%     \item Explainable Artificial Intelligence Applications in Cyber Security: State-of-the-Art in Research \cite{zhang2022explainable}
%     \item An Explainable Machine Learning Framework for Intrusion Detection Systems \cite{wang2020explainable}
% \end{itemize}、

% \textbf{Applications}
% \begin{itemize}
%     \item Explaining anomalies detected by autoencoders using Shapley Additive Explanations \cite{antwarg2021explaining}
% \end{itemize}
%%%%%%%%%%%%%%%%%%

\section{Methodologies}\label{sec:method}
Machine learning models are also regarded as "black box" models because of the difficulty to explain and interpret. But modern interpretability in machine learning has been improved and implemented from complex ML models, which can be categorized into two kinds of interpretability: global and local interpretability. Global interpretability aims to understand the model structure based on its features, while local interpretability wants to find out the reason for making the decision. SHAP is a method that can be offered global and local interpretability of the model. In this paper, SHAP interpretability can be used to extract the most contributed features as a method of feature selection and also explained the model. SHAP can be introduced in the following selection. SHAP commonly connects to the Shapley value, thus, the Shapley value also can be depicted in this section.

\subsection{SHAPLEY value}
The Shapley value introduced by Shapely~\cite{shapley201617}, is a method from cooperative game theory to determine individual contributions among the features. The method can be contributed to machine learning prediction, weighted, and encapsulated through the contribution of each feature value. The Shapley value of a feature value is its average contribution to the prediction in different coalitions:

\begin{equation}\label{eq:phi}
\small
	\phi_j(\nu) = \sum_{S\subseteq \{x_1,...,x_p\} \backslash \{x_j\}} \frac{\left|S\right|!(p-\left|S\right|-1)!}{p!}(\nu(S\cup\{x_j\}) - \nu(S))
\end{equation}

where:
\begin{itemize}
	\item $S\subseteq \{x_1,...,x_p\} \backslash \{x_j\}$ represents that S is a subset of $p$ features in the model, x is the vector of feature value of the instance to be explained, and p is the number of features.
	\item $\nu(S)$ is the prediction for feature values in set $S$.
\end{itemize}

The Shapley value is the sole attribute technique that satisfies four properties: Efficiency, Dummy, Symmetry, and Additivity. These four properties are required for a definition of a fair prediction (payout).

\textbf{Efficiency:} The feature contributions have to add up to the difference of prediction for the average predicted value.

\textbf{Dummy:} The feature $j$ does not modify the predicted value regardless of which irrespective of feature values it is added to, this should have a 0 Shapley value.

\begin{equation}
\small
	\begin{aligned}
		\nu(S\cup\{x_j\}) = \nu(S), for\: all \: S \subseteq \{x_1,...x_p\},\: then\: \phi_j = 0
	\end{aligned}
\end{equation}

\textbf{Symmetry:}If the two features $j$ and $i$ contribute equally to all possible coalition, then the contributions of these two features should be the same.
\begin{equation}
\small
	\begin{aligned}
		\nu(S\cup\{x_j\}) = \nu(S\cup{x_i}),\: for\: all\: S \subseteq \{x_1,...,x_p\} \backslash \{x_j\}, \\ 
   then\: \nu_j = \nu_i
	\end{aligned}
\end{equation}

\textbf{Additivity:} The respective Shapley values for a game with combined payouts $\nu$ + $\nu^+$, which can be show as follows: 

\begin{equation}
	\phi_j + \phi_j^+
\end{equation}

\subsection{Shapley Additive Explanations (SHAP)}

SHAP proposed by Lundberg LEE~\cite{lundberg2017unified}, is a unified framework for the interpretation prediction of models. SHAP aims to explain the prediction of an instance $x$ by computing the contribution of each feature to the final predictions of the model. This contribution of each feature can be either positive or negative. One innovation of the SHAP explanation method brings to the table is that the shapley value explanation is represented as a linear model - an additive feature attribution method. The strength of SHAP can be computed for any model rather than only a linear model. Moreover, instead of only explaining local interpretations, the SHAP interprets global interpretation by summing the attributed value of each input feature and averaging features individually. SHAP defines the explanation of an instance $x$ can be expressed as follows:

\begin{equation}\label{eq:shap}
	g(z') = \phi{_0} + \sum_{j=1}^{M} \phi_j z'{_j}
\end{equation}

where
\begin{itemize}
	\item $g$ represents the explanation model.
	\item $z'$ represents simplified features, and $z' \in \{0,1\}^M$.
	\item $M$ represents the maximum coalition size.
	\item $\phi{_j} \in R$  is the shapley value, representing the feature attribution for a feature $j$.
\end{itemize}

\begin{figure*}[t]
	\centering
	\includegraphics[width=0.7\linewidth]{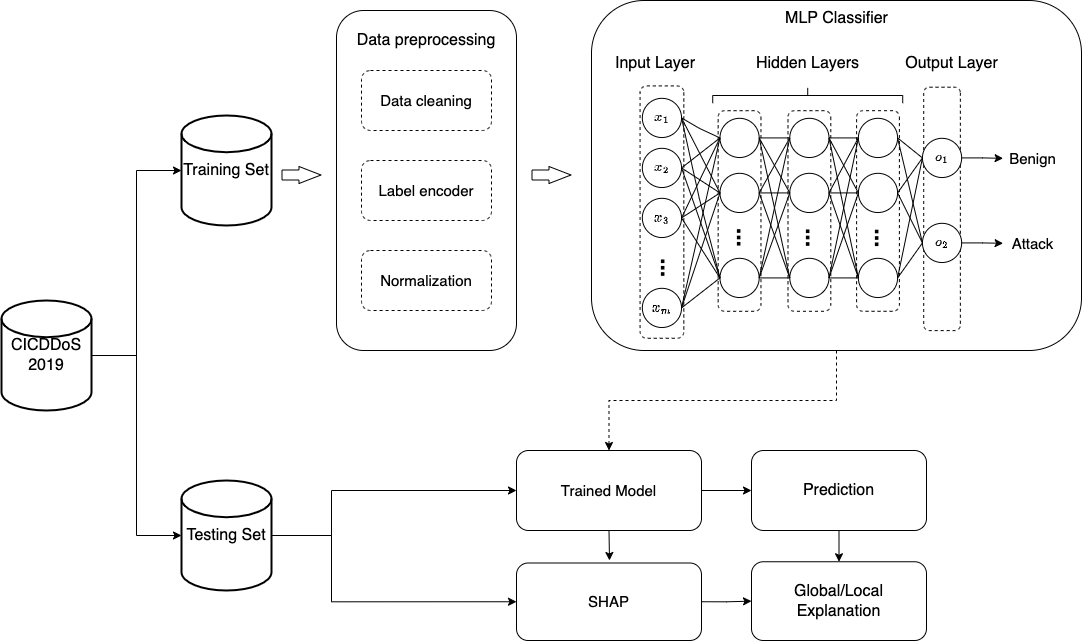}
	\caption{Overview of our proposed framework}
	\label{alg:mlp}
\end{figure*}

The SHAP scores offer two significant advantages over other methods. First, SHAP has a solid theoretical foundation in game theory. Shapley scores are the only solutions that satisfy four properties: Efficiency, Symmetry, Dummy, and additivity. It helps to unify the field of interpretable machine learning. 
% Second, SHAP has a fast computation for machine learning models compared to computing Shapley value directly. 
%\Tui{This paragraph needs to be rewritten}
%SHAP can also satisfy these since it obtains Shapley values from linear models. Second, SHAP combines LIME and Shapley values.

\subsection{Proposed Methodology}
The proposed framework is divided into the classification, and model explanation. 

% \subsubsection{Feature Selection}
Before feeding the data into the classifier, selecting the most significant important, and relevant subset of features is an important step in the optimization process that can improve the performance efficiency of the model. Feature selection is an important phase to select the most important features, which can 1)reduce feature space or dimensional; 2) reduce the computational time; 3) improve the model efficiency. Several feature selection techniques are depicted in~\cite{gebreyesus2023machine} to gain the most relevant features, including Random Forest with SHAP-based, and XGB with Permutation-based, SHAP-based, etc. Therefore, in this work, we adopt a combination method of XBGoost-based feature importance, Permutation importance, and SHAP feature importance to select the top 20 important features to feed into the classifier to classify the benign traffic and attacks. Table~\ref{table:Feature} shows the top 20 important features based on XGB-based feature importance methods. According to the occurrence frequency of each feature, we select the top 20 features based on the order of most occurrence frequency from high to low. As a result, the top 20 most important features in terms of single feature importance (FI) and multiple feature importance are shown in Table~\ref{table:Feature}. Note that in Table~\ref{table:Feature}, three of the top 20 XGB-based feature importance methods are based on counting single features. The top 20 important features of each XGB-based technique shown in the figure are based on counting single features. Similarly, for the single-multiple importance features, the top 20 most important features of each XGB-based technique of each attack were counted separately, and then the top 20 were selected from the frequency of occurrence of each feature. For example, we counted the top 20 important features of each DNS, SNMP, LDAP, and NetBIOS attack separately, and then selected the top 20 based on the frequency of occurrence of each feature of each attack. As a consequence, the top 20 important features of one-multiple feature importance are shown in Table~\ref{table:Feature}.

% In the classification, the top n important features are selected separately based on XGB-based feature importance techniques, including XGBoost-based feature importance, Permutation importance, and SHAP feature importance. After selecting the top 20 important features of each attack type separately, we combine them and select the top n important features by counting the most occurrences of the top n features from each attack type. 

After feature selection, we use the MLP classifier to classify the benign traffic and the attacks. Note that the benign traffic is classified as 1, while the attack/malicious is classified as 0. The Multi-Layer Perceptron (MLP) is a feed-forward network, it has an input layer, multiple hidden layers, and an output layer (equal to attacks and benign traffic). The setting of hyperparameters in this proposed framework is employed to better classify benign traffic and attacks, which is based on the MLP classifier from\cite{wei2021ae}. We have experimented with the best optimized MLP architecture the one that uses 5 layers – 1 input layer, 3 hidden layers, and 1 output layer. The Hyperparameters of hidden layer size are [23, 15, 10]. Note that the output layer mostly uses a sigmoid function for binary class classification problems. 

After training the classifier model of MLP, then we obtain a trained classifier model (also refer a black box model), and the prediction results. Finally, we use the SHAP explanation technique to explain the model based on the global and local explanation. The overview of our the proposed framework of classification is depicted in Fig.~\ref{alg:mlp}. To explain the trained black box model, we use Kernel SHAP to obtain the global and local explanation (Fig.~\ref{alg:mlp}). Kernel SHAP is suitable for the classification model with tabular data, which can be applied to any model. 

\begin{table*}[]
  \centering
  \setlength\tabcolsep{4.5pt}\renewcommand\arraystretch{1.25}
  \caption{Top 20 Important Features Based on Three XGB-based Feature Importance Methods}
  \label{table:Feature}
  \begin{tabular}{ccccc|c}
  \midrule
     & XGBoost Feature Importance & Permutation Importance & SHAP Feature Importance & one-one-FI & one-multiple-FI\\ \hline
     1 & Bwd Packets/s & Inbound & Min Packet Length & Protocol & Protocol \\
     2 & Protocol & Protocol & Fwd Packet Length Mean & URG Flag Count & URG Flag Count \\
     3 & Inbound & Init Win bytes forward & Fwd Packet Length Min & Flow Duration & Init Win bytes forward \\
     4 & Init Win bytes backward & URG Flag Count & Average Packet Size & Init Win bytes forward & Min Packet Length \\
     5 & URG Flag Count & Min Packet Length & Fwd Packet Length Max & Fwd Packet Length Min & Fwd Packet Length Max \\
     6 & ACK Flag Count & Fwd Packet Length Max & Max Packet Length & Min Packet Length & Inbound \\
     7 & Total Fwd Packets & Bwd Packets/s & Protocol & Fwd Packets/s & Total Backward Packets \\
     8 & Flow Duration & Max Packet Length & Flow IAT Mean & Max Packet Length & Flow IAT Min \\
     9 & Init Win bytes forward & ACK Flag Count & Init Win bytes forward & Flow IAT Min & Flow Duration \\
     10 & Fwd Packet Length Min & Fwd Packet Length Min & URG Flag Count & Fwd Packet Length Max & Bwd Packets/s \\
     11 & Bwd IAT Total & Flow Duration & Flow Duration & Average Packet Size & Fwd Packets/s \\
     12 & Total Backward Packets & Average Packet Size & Flow IAT Std & Bwd Packets/s & Max Packet Length \\
     13 & Packet Length Std & Init Win bytes backward & Flow IAT Min & Inbound & Init Win bytes backward \\
     14 & Min Packet Length & Fwd Packets/s & Fwd Packets/s & Init Win bytes backward & Fwd Packet Length Min \\
     15 & Active Min & Fwd IAT Min & Total Length of Fwd Packets & ACK Flag Count & Fwd Packet Length Mean \\
     16 & Fwd Packets/s & Packet Length Mean & Fwd IAT Total & Total Fwd Packets & Average Packet Size \\
     17 & Max Packet Length & Flow IAT Min & Total Fwd Packets & Total Backward Packets & Packet Length Std \\
     18 & Flow IAT Min & Flow IAT Mean & Total Backward Packets & Flow IAT Mean & Packet Length Mean \\
     19 & Fwd Packet Length Max & SYN Flag Count & act data pkt fwd & Packet Length Std & SYN Flag Count \\
     20 & Average Packet Size & CWE Flag Count & Fwd IAT Mean & Bwd IAT Total & Bwd IAT Total \\
     \midrule
  \end{tabular}
\end{table*}

\section{Data and Experiment Setup}\label{sec:re}

%\subsection{Experiment Setup}
Our experiments were carried out using the system setup shown in Table~\ref{table:Mat}. 

\begin{table}[h]
	\centering
	\footnotesize
	\caption{Implementation environment specification}
	\label{table:Mat}
	\begin{tabular}{p{2.6cm} | p{3.8cm}}
		\hline
		\textbf{Unit}   & \textbf{Description}\\ \hline
		Processor   & 3.4GH$_z$  Inter Core i5 \\ \hline
		RAM  &  16GB      \\ \hline
		OS  &  MacOS Big Sur   11.4  \\ \hline	
		Packages used  &  tensorflow 2.0.0, sklearn 0.24.1    \\ \hline	
	\end{tabular}
\end{table}

\begin{table} [h]
	\setlength{\tabcolsep}{4.5mm}
	\caption{The number of records in CICDDoS2019 }
	\label{table:no_dataset}
	\begin{tabular}{cccc}
		\midrule 
		{\textbf{dataset}} & {\textbf{total}} & {\textbf{benign}} & {\textbf{Attacks}} \\ 
		\midrule
            \addlinespace[0.5ex]
		Training day & 50,063,112	& 56,863	& 50,006,249 \\ 
            \addlinespace[0.5ex]
		Testing day & 20,364,525    & 56,965    & 20,307,560 \\ 
		\midrule
	\end{tabular}
\end{table}

\subsection{Data Pre-prosessing}
In this section, we discuss the methodologies we used to process our dataset in order to feed it into our proposed model.

\subsubsection{CICDDOS2019 Dataset}
In this study, we use CICDDoS2019~\cite{sharafaldin2019developing} dataset that has been widely used for DDoS attack detection and classification. The dataset contains a large amount of up-to-date realistic DDoS attack samples as well as benign samples. The total number of records contained in CICDDoS2019 is depicted in Table~\ref{table:no_dataset}. Table~\ref{table:no_dataset} illustrated that all data was captured in two days: training day and Testing day. Furthermore, all attack types can be collected from the application layer by using TCP/UDP-based protocols, which can be separated as reflection and exploitation-based DDoS attacks. The taxonomy of the CICDDoS2019 dataset is depicted in Table~\ref{table:ddos}, including 13 different attack types, in terms of MSSQL, SSDP, DNS, LDAP, NetBIOS, SNMP, NTP, TFTP, UDP, UDP-Lag, SYN, PORTMAP, WebDDoS, and its attack time. Note that the "WebDDoS" attack was collected and saved together with the "UDPLag" attack file. In this research, In this research, we use four reflection-based attack types: DNS, LDAP, NetBIOS, and SNMP for classification and explanation, which is captured from 10:52 to 12:23 for the training day on January 12th. 
%\begin{center}

% \begin{table*}[]
%    {\renewcommand{\arraystretch}{1.5}
%    \centering
%    \caption{The Taxonomy of CICDDOS2019 Dataset}
%    \label{table:no_dataset}
%    \begin{tabular}{c|c|ccc}  
%    \midrule
%    Collect Day & DDOS Attack & Protocol & Attack types & Attack time \\ \hline
%    \multirow{11}{*}{Training day} & \multirow{8}{*}{Reflection-based} & \multirow{2}{*}{TCP-based} & MSSQL & 11:36-11:45 \\ %\cline{4-5} 
%    &  &  & SSDP & 12:27-12:37 \\ \cline{3-5} 
%    &  & \multirow{4}{*}{TCP/UDP-based} & DNS & 10:52-11:05 \\ 
%    &  &  & LDAP & 11:22-11:32 \\
%    &  &  & NETBIOS & 11:50-12:00 \\
%    &  &  & SNMP & 12:12-12:23 \\ \cline{3-5} 
%    &  & \multirow{2}{*}{UDP-based} & NTP & 10:35-10:45 \\
%    &  &  & TFTP & 13:35-17:15 \\ \cline{2-5}
%    & \multirow{3}{*}{Exploitation-based} & \multirow{2}{*}{UDP-based} & UDP & 12:45-13:09 \\
%    &  &  & UDP-Lag & 13:11-13:15 \\ \cline{3-5}
%    &  & TCP-based & SYN & 13:29-13:34 \\
%    \hline
%    \multirow{7}{*}{Testing day} & \multirow{4}{*}{Reflection-based} & \multirow{3}{*}{TCP/UDP-based} & PortMap & 09:43-09:51 \\
%    &  &  & NetBIOS & 10:00-10:09 \\
%    &  &  & LDAP & 10:21-10:30 \\ \cline{3-5}
%    &  & TCP-based & MSSQL & 10:33-10:42 \\ \cline{2-5}
%    & \multirow{3}{*}{Exploitation-based} & \multirow{2}{*}{UDP-based} & UDP & 10:53-11:03 \\
%    &  &  & UDP-Lag & 11:14-11:24 \\ \cline{3-5}
%    &  & TCP-based & SYN & 11:28-17:35 \\
%    \midrule
%    \end{tabular}
% \end{table*}

\begin{table}[]
   \centering
   \caption{The Taxonomy of CICDDOS2019 Dataset}
   \label{table:ddos}
   {\renewcommand{\arraystretch}{1.5}
   \begin{tabular}{c|c|ccc}  
   \midrule
   \begin{tabular}[c]{@{}c@{}}Collected \\ Day\end{tabular} & \begin{tabular}[c]{@{}c@{}}DDOS \\ 
    Attack\end{tabular} & Protocol & \begin{tabular}[c]{@{}c@{}}Attack \\ Types\end{tabular} & \begin{tabular}[c]{@{}c@{}}Attack \\ Time\end{tabular} \\ \hline
        \multirow{11}{*}{\begin{tabular}[c]{@{}c@{}}Training \\ day\end{tabular}} & \multirow{8}{*}{\begin{tabular}[c]{@{}c@{}}Reflection-\\ based\end{tabular}} & \multirow{2}{*}{TCP} & MSSQL & 11:36-11:45 \\ %\cline{4-5} 
        &  &  & SSDP & 12:27-12:37 \\ \cline{3-5} 
        &  & \multirow{4}{*}{\textbf{TCP/UDP}} & \textbf{DNS} & 10:52-11:05 \\ 
        &  &  & \textbf{LDAP} & 11:22-11:32 \\ 
        &  &  & \textbf{NetBIOS} & 11:50-12:00 \\ 
        &  &  & \textbf{SNMP} & 12:12-12:23 \\ \cline{3-5} 
        &  & \multirow{2}{*}{UDP} & NTP & 10:35-10:45 \\ 
        &  &  & TFTP & 13:35-17:15 \\ \cline{2-5} 
        & \multirow{3}{*}{\begin{tabular}[c]{@{}c@{}}Exploitation-\\ based\end{tabular}} & \multirow{2}{*}{UDP} & UDP & 12:45-13:09 \\ 
        &  &  & UDP-Lag & 13:11-13:15 \\ \cline{3-5} 
        &  & TCP & SYN & 13:29-13:34 \\ 
        \midrule
        \multirow{7}{*}{\begin{tabular}[c]{@{}c@{}}Testing \\ day\end{tabular}} & \multirow{4}{*}{\begin{tabular}[c]{@{}c@{}}Reflection-\\ based\end{tabular}} & \multirow{3}{*}{TCP/UDP} & PortMap & 09:43-09:51 \\ 
        &  &  & NetBIOS & 10:00-10:09 \\ 
        &  &  & LDAP & 10:21-10:30 \\ \cline{3-5} 
        &  & TCP & MSSQL & 10:33-10:42 \\ \cline{2-5} 
        & \multirow{3}{*}{\begin{tabular}[c]{@{}c@{}}Exploitation-\\ based\end{tabular}} & \multirow{2}{*}{UDP} & UDP & 10:53-11:03 \\ 
        &  &  & UDP-Lag & 11:14-11:24 \\ \cline{3-5} 
        &  & TCP & SYN & 11:28-17:35 \\ 
        \midrule
   \end{tabular}}
\end{table}

\subsubsection{data imbalance}
In classification scenarios, Solving data imbalance is a common issue in a kind of machine learning problem. Table~\ref{table:no_dataset} shows the number of records in the CICDDoS2019 dataset, which can be seen as having an uneven distribution of the number of records. For example,  benign samples have a lower number of records (56,863), while attacks have a large number of records (50,006,249) on training day collection. In this research, we use a DSN attack as a one-to-one classification, including benign traffic (labeled as 1) and a DNS attack (labeled as 0), and a one-to-multiple classification, including benign traffic (labeled as 1) and four attack types: DNS, LDAP, NetBIOS, SNMP attacks (labeled as 0). In both two scenarios, all benign traffic has been extracted first, then combined with 0.1\% four attacks separately as the classification and explanation dataset. 

\subsubsection{Data Cleaning}
The original CICDDoS2019 dataset contained 88 features. As suggested by \cite{almiani2021ddos}, we also removed the features not contributing to detecting DDoS attacks. These include the feature such as "Unnamed", "Flow ID", "Source IP", "Destination IP", "Source Port", "Destination Port", "Timestamp", "Flow Bytes", "Flow Packets", and "SimilarHTTP". After the exclusion of these 10 features, we have 78 features to work with. Following the recommendation of the work by \cite{samom2021distributed, batchu2022integrated}, we cleaned up the values containing NaN (not a number), redundant, and infinity values, which can significantly affect the data modeling efficiency and data knowledge discovery. 

\subsubsection{Label Encoding}
We substituted the categorical labels with deep models as they only operate on float/numeric values. LabelEncoding is one approach to achieve this, where each unique category is assigned a numeric value. After employing LabelEncoding, we have obtained the label of the DDoS traffic flow as either benign traffic (1) or malicious traffic (0). Note that a one-to-one scenario represents benign and malicious (DNS attack traffic), while a one-to-all scenario represents benign and malicious (including DNS, LDAP, NetBIOS, and SNMP four attack types and labeled as malicious).

\subsubsection{Data Normalization}
The CICDDoS2019 datasets contain some features with very high variance in terms of value between the minimum and the maximum (e.g., "Min Packet Length", "Flow Duration", "URG Flag Count", "Fwd Packet Length Mean", etc.). We applied a normalization strategy to eliminate the impacts of big variance of the values across the features thus reducing the execution time for model training and improving accuracy.
There are several widely used methods to perform feature scaling, including Z Score, standardization, and normalization. As proposed by \cite{ur2021diddos}, we use MinMax-based normalization for our feature scaling. This method maps the original range of each feature into a new range with  Equation~(\ref{eq:normalised})

\begin{equation}\label{eq:normalised}
Z_i = \frac{Z_i - min}{max -  min}
\end{equation}

where Z$_i$ donates all the normalized numeric values ranging between [0-1]; $max$ and $min$ donate the maximum and minimum values from all data points.

\subsection{Performance Matrix}
To evaluate the performance of our model, we used the following metrics: classification accuracy, precision, recall, and F1 score. Table \ref{table:Matrix} illustrates the confusion matrix.
\begin{table}[h]
	\centering
	\caption{Confusion Matrix}
	\label{table:Matrix}
	\begin{tabular}{| p{2.8cm} | c | p{1.5cm} | p{1.5cm} |}
		\hline
		\multicolumn{2}{|c|}{ \multirow{2}{*}{Total Population} } &   \multicolumn{2}{c|}{Predicted Condition} \\
		\cline{3-4}
		\multicolumn{2}{|c|}{} & Normal & Anomaly \\
		\hline
		\multirow{2}{*}{Actual Condition} & Normal & TN & FP \\
		\cline{2-4}
		&Anomaly & FN & TP \\
		\hline
	\end{tabular}
\end{table}

where;
\begin{itemize}
	\item True Positive (TP) indicates an anomalous data point correctly classified as anomalous. 
	\item True Negative (TN) indicates a normal data point correctly classified as normal.
	\item False Positive (FP) indicates a normal data point incorrectly classified as anomalous.
	\item False Negative (FN) indicates an anomalous data point incorrectly classified as normal.
\end{itemize}

Based on the aforementioned terms, the evaluation metrics are calculated as follows: \\
\begin{itemize}
    \item True Positive Rate (also known as Recall) estimates the ratio of the correctly predicted samples of the class to the overall number of instances of the same class.
        \begin{equation}\label{eq:TPR}
	    TPR (Recall) = \frac{TP}{TP + FN}
        \end{equation}

    \item False Positive Rate (FPR) presents the proportion of data points correctly classified as anomalous.
        \begin{equation}\label{eq:FPR}
	    FPR = \frac{FP}{FP + TN}
        \end{equation}
    \item Precision (Pre) measures the quality of the correct predictions.
        \begin{equation}\label{eq:PPV}
	    Precision = \frac{TP}{TP + FP}
        \end{equation}
    \item F1-Score computes the trade-off between precision and recall.
        \begin{equation}\label{eq:F-measure}
	    F1-score = 2\times\left(\frac{Precision\times Recall}{Precision + Recall}\right)
        \end{equation}
    \item Accuracy (Acc) measures the total number of data samples correctly classified.
        \begin{equation}\label{eq:ACC}
    	Accuracy = \frac{TP+TN}{TP + TN + FP + FN}
        \end{equation}
\end{itemize}

The Area Under the Curve (AUC) computes the area under the Receiver Operating Characteristics (ROC) curve which is plotted using the true positive rate on the y-axis and the false positive rate on the x-axis over different thresholds. Mathematically, the AUC is computed as shown in Equation~(\ref{eq:auc}).

\begin{equation}\label{eq:auc}
	AUC_{ROC}=\int_{0}^{1} \frac{TP}{TP+FN}d\frac{FP}{TN+FP}
\end{equation}

% \section{results and Evaluations} 

% Fig~\ref{alg:shap}. is the overview of SHAP usage to interpret of model's prediction. This explanation aims to describe which features contribute to predicting the target.

% \begin{figure}[h!]
% 	\centering
% 	\includegraphics[width=0.7\linewidth]{Figures/SHAP_interpret.png}
% 	\caption{Overview of SHAP  Usage to Interpret of Model's Prediction}
% 	\label{alg:shap}
% \end{figure}

% Add:\textit{table1: Hyperparameters}
% \textit{table2/Figure: Top-n important features obtained from SHAP/xgb}

\section{Experiment Results and Evaluations}\label{sec:Experiments}
The experiment results are evaluated based on the performance matrix, in terms of accuracy, precision, recall, and F1-score. The confusion matrix of accuracy refers to the degree of closeness of a given set of traffic samples and its true samples, while F1-score is evaluated the classifier performance. Furthermore, we also give an explanation of the decision-making of the classifier model based on global and local explanations.

% \begin{figure}
% \includegraphics[width=8cm]{Figures/one-one.png}\\
% \vspace{-0.5cm} 
% \includegraphics[width=8cm]{Figures/Global-one.png} 
% \end{figure}

\subsection{DDoS Binary Classification}
Table~\ref{table: Matrix} shows the experimental results of the proposed model, which achieved over 99\% of all performance metrics in terms of over 99.91\% of all precision, 99.83\% recall, and 99.87\% F1 score for benign traffic, while attacks achieved over 99.80\% of all precision, 99.90\% recall, and 99.84\% F1 score. Furthermore, the feature selection results achieved better performance in terms of precision, recall, and F1 score in both one-to-one and one-to-all scenarios than the performance of using all features. Moreover, Table~\ref{table: Total_Matrix} shows that the total performance of accuracy of feature selection performed better than the accuracy results of using all features. For example, one-to-one and one-to-all scenarios with feature selection achieved 99.95\% accuracy, while one-to-one and one-to-all scenarios with all features achieved 99.86\% and 99.91\% accuracy separately. 

\begin{table*}[]
    \centering
    \caption{Performance Matrix based on Benign and malicious}
    \label{table: Matrix}
    {\renewcommand{\arraystretch}{1.5}
    \begin{tabular}{ccccccccccccc}
    \midrule
    \multirow{2}{*}{\begin{tabular}[c]{@{}c@{}}Malicious\\ Type\end{tabular}} & \multicolumn{3}{c}{One-to-one} & \multicolumn{3}{c}{\textbf{One-to-one-FS}} & \multicolumn{3}{c}{One-to-all} & \multicolumn{3}{c}{\textbf{One-to-all-FS}} \\ 
    & Precision & Recall & F1-score & Precision & Recall & F1-score & Precision & Recall & F1-score & Precision & Recall & F1-score \\ \hline
    Benign & 99.91 & 99.83 & 99.87 & 100 & 99.91 & 99.95 & 99.93 & 99.81 & 99.87 & 100 & 99.91 & 99.95 \\
    Attack & 99.80 & 99.90 & 99.85 & 99.90 & 100 & 99.95 & 99.90 & 99.96 & 99.93 & 99.90 & 100 & 99.95 \\
    \midrule
    \end{tabular}}
\end{table*}

% \begin{table}[h]
%     \centering
%     \caption{Performance Matrix based on Benign and Attack}
%     \label{table:Matrix}
%     {\renewcommand{\arraystretch}{1.5}
%     \begin{tabular}{ccccccc}
%     \midrule
%          \multirow{2}{*}{\begin{tabular}[c]{@{}c@{}}DDoS\\ Attack\end{tabular}} & \multicolumn{3}{c}{one-to-one} & \multicolumn{3}{c}{one-to-multiple} \\ 
%          & Pre & Re & F1 & Pre & Re & F1 \\ \hline
%          Benign & 99.85 & 99.70 & 99.78 & 99.87 & 99.75 & 99.81 \\
%          Attack & 99.80 & 99.90 & 99.85 & 99.87 & 99.93 & 99.90 \\
%     \midrule
%     \end{tabular}}
% \end{table}

\begin{table}[h]
    \centering
    \caption{Total Performance Matrix}
    \label{table: Total_Matrix}
    {\renewcommand{\arraystretch}{1.5}
    \begin{tabular}{cccccc}
    \midrule
    Performance & Acc & Pre & Re & F1 & AUC \\ \hline
    One-to-one & 99.86 & 99.91 & 99.83 & 99.87 & 99.99 \\
    \textbf{One-to-one-FS} & 99.95 & 99.90 & 100 & 99.95 & 100 \\
    One-to-all & 99.91 & 99.93 & 99.81 & 99.87 & 99.95 \\
    \textbf{One-to-all-FS} & 99.95 & 100 & 99.91 & 99.95 & 99.99 \\
    \midrule
    \end{tabular}}
\end{table}

% \begin{table}[h]
%     \centering
%     \caption{Total Performance Matrix}
%     \label{table:Total_Matrix}
%     {\renewcommand{\arraystretch}{1.5}
%     \begin{tabular}{cclcccc}
%     \midrule
%     Total Performance & \multicolumn{2}{c}{Acc} & Pre & Re & F1 & AUC \\ \hline
%     one-to-one & \multicolumn{2}{c}{99.82} & 99.80 & 99.90 & 99.85 & 99.99 \\
%     one-to-multiple & \multicolumn{2}{c}{99.87} & 99.87 & 99.75 & 99.81 & 99.96\\
%     \midrule
%     \end{tabular}}
% \end{table}
\subsection{Model Explanation}
In order to establish the trustworthiness of the prediction results of the proposed model, we use SHAP to explain the decision-making of the prediction results and to explain why DDoS traffic flows are classified as benign traffic or attacks. As mentioned above, SHAP is a method to generate an interpretation of the classifier's prediction based on Shapley values and the most important features. The explanation of SHAP can: 1) represent the importance of the features based on the SHAP values and presented in decreasing order from high to low; 2) provide insight into the contribution of high or low feature values to the prediction outcomes. We employ SHAP to investigate how the classifier correctly makes the classification decision between benign and malicious traffic by global and local explanations. 

\subsubsection{Global Explanation}
\begin{figure*}[t]%
    \centering    
    \subfloat[\label{fig: OAll-mean} \centering Mean Plot with All Features]{{\includegraphics[width=7cm]{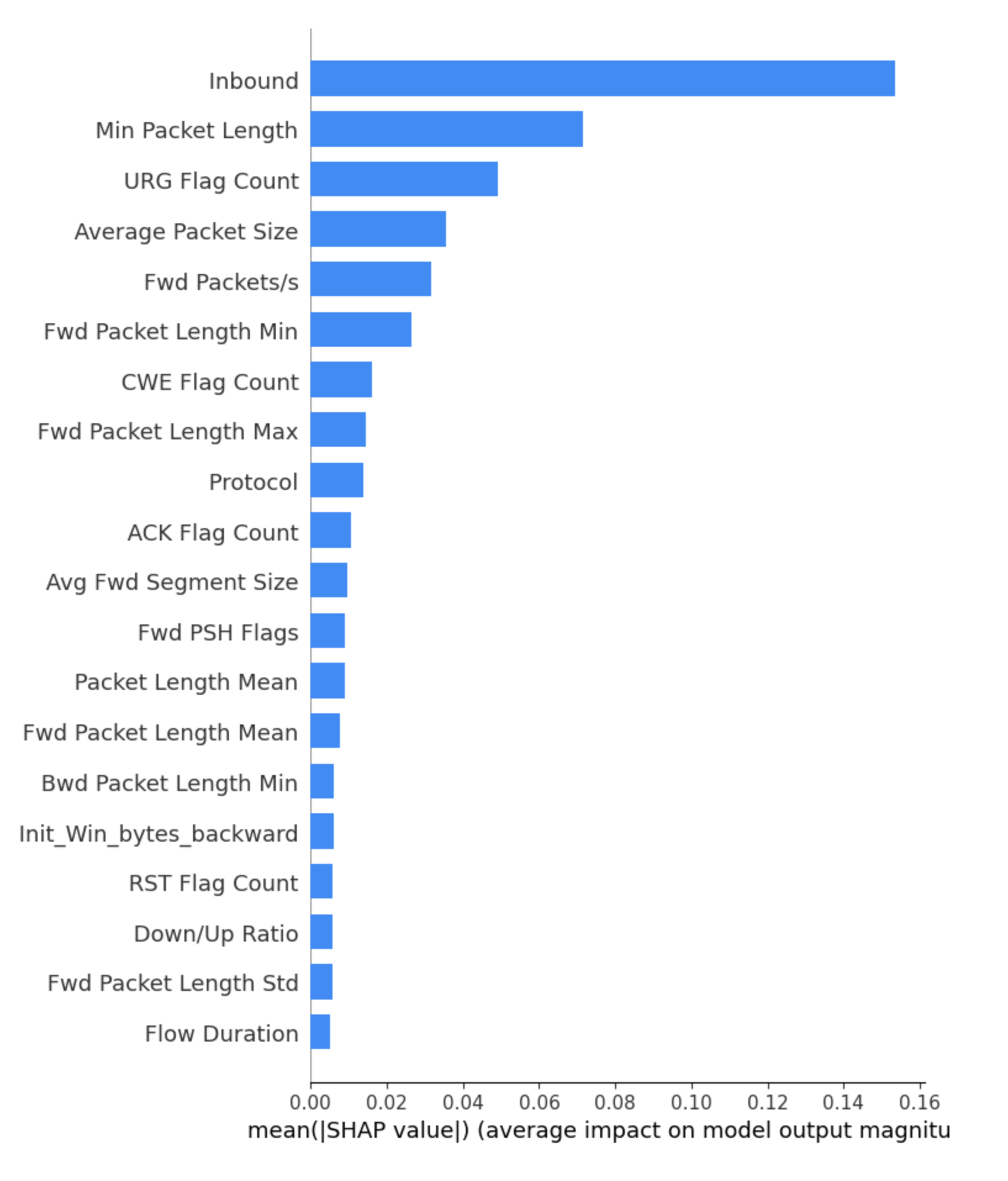} }}%
    \subfloat[\label{fig: OAll-summary} \centering Summary Plot with All Features]{{\includegraphics[width=7cm]{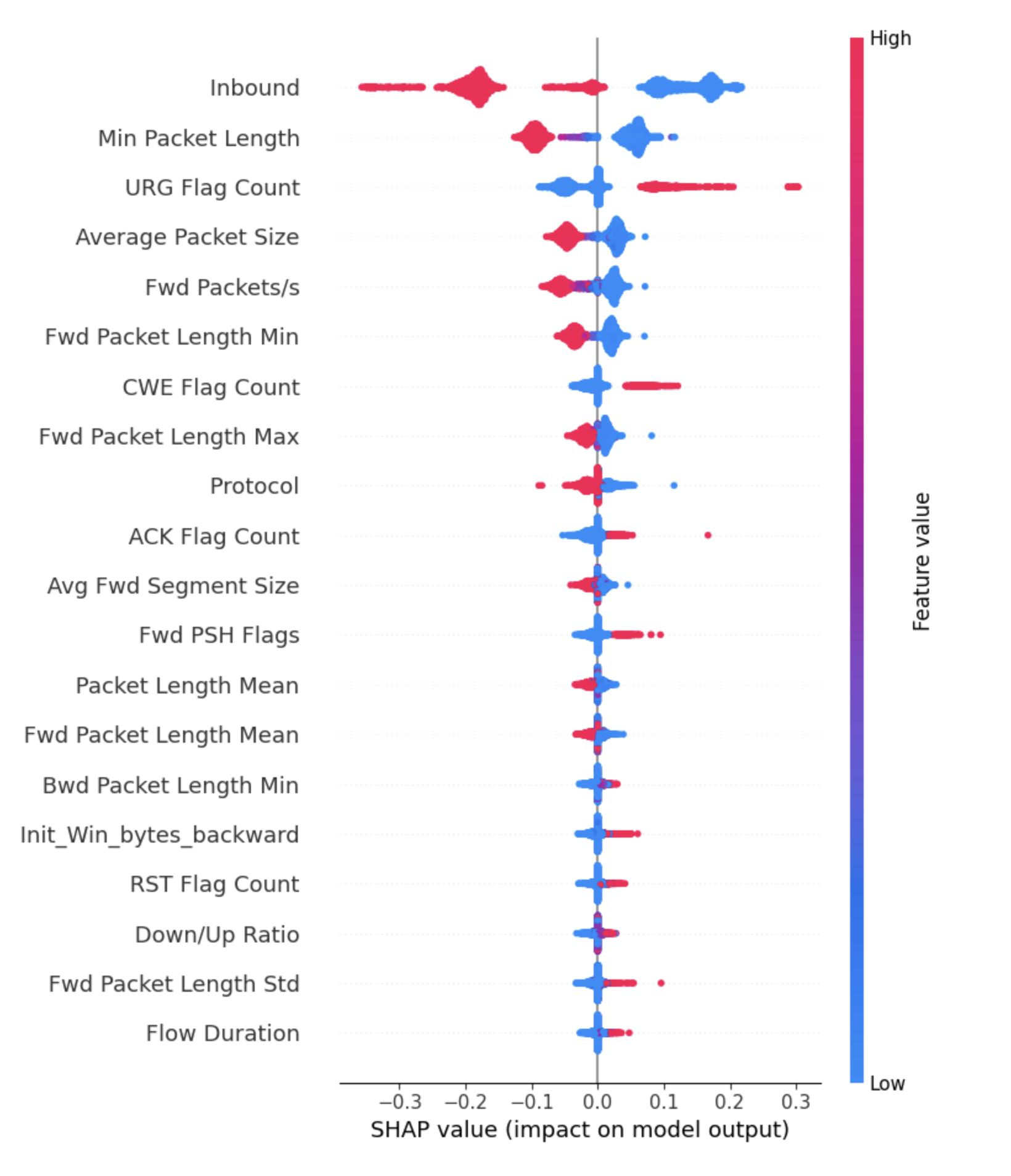} }}%% \par
    \par
    \subfloat[\label{fig: OFS-mean} \centering Mean Plot with Selected Features]{{\includegraphics[width=7cm]{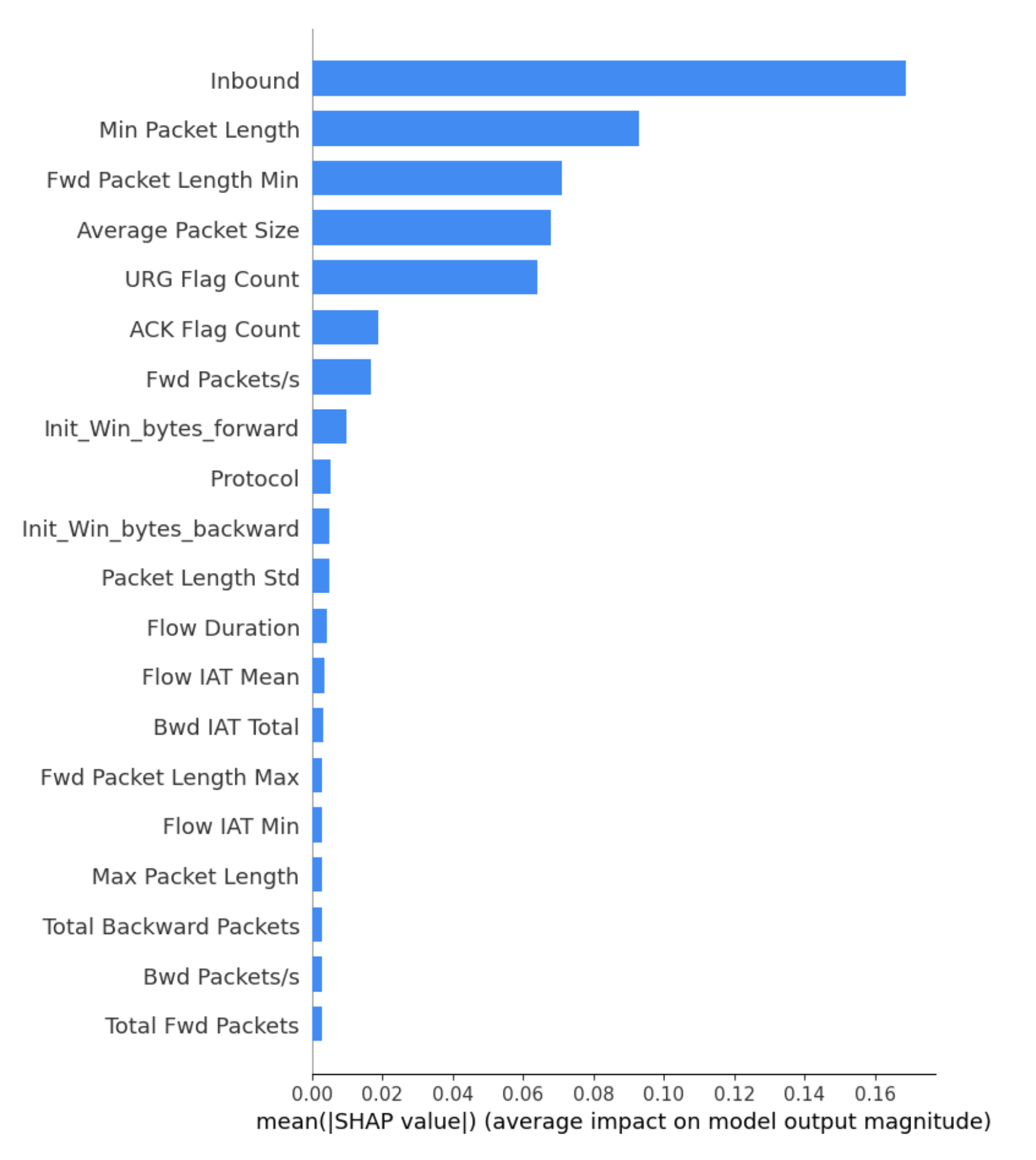} }}%
    \subfloat[\label{fig: OFS-summary} \centering Summary Plot with Selected Features]{{\includegraphics[width=7cm]{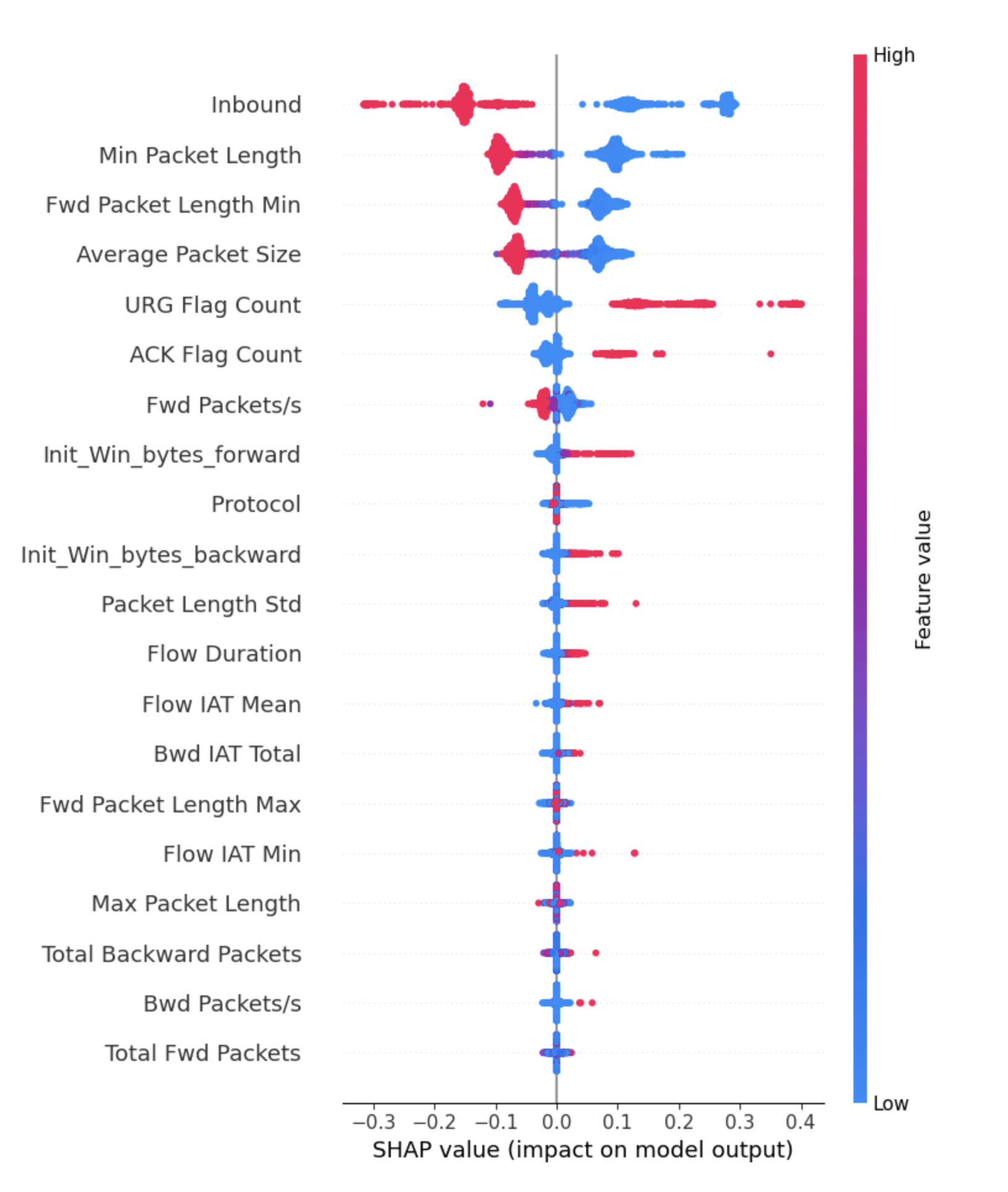} }}%
    \caption{One-to-one (Benign vs. DNS): the Explanation of SHAP Values}
    \label{fig: glo-One}%
\end{figure*}

\begin{figure*}[t]%
    \centering
    \subfloat[\label{fig: MAll-mean}\centering Mean Plot with All Features]{{\includegraphics[width=7cm]{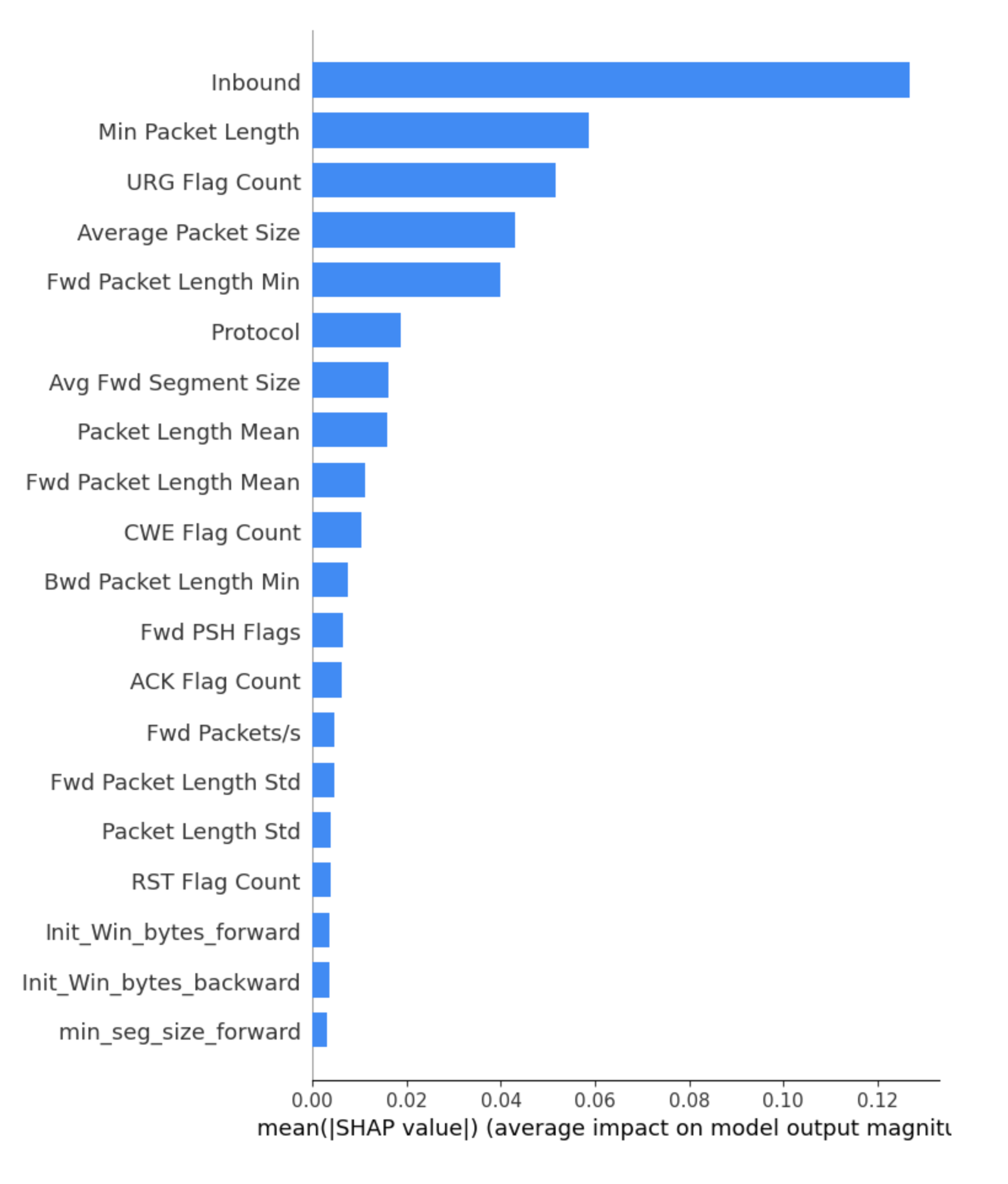} }}%    
    \subfloat[\label{fig: MAll-summary}\centering Summary Plot with All Features]{{\includegraphics[width=7cm]{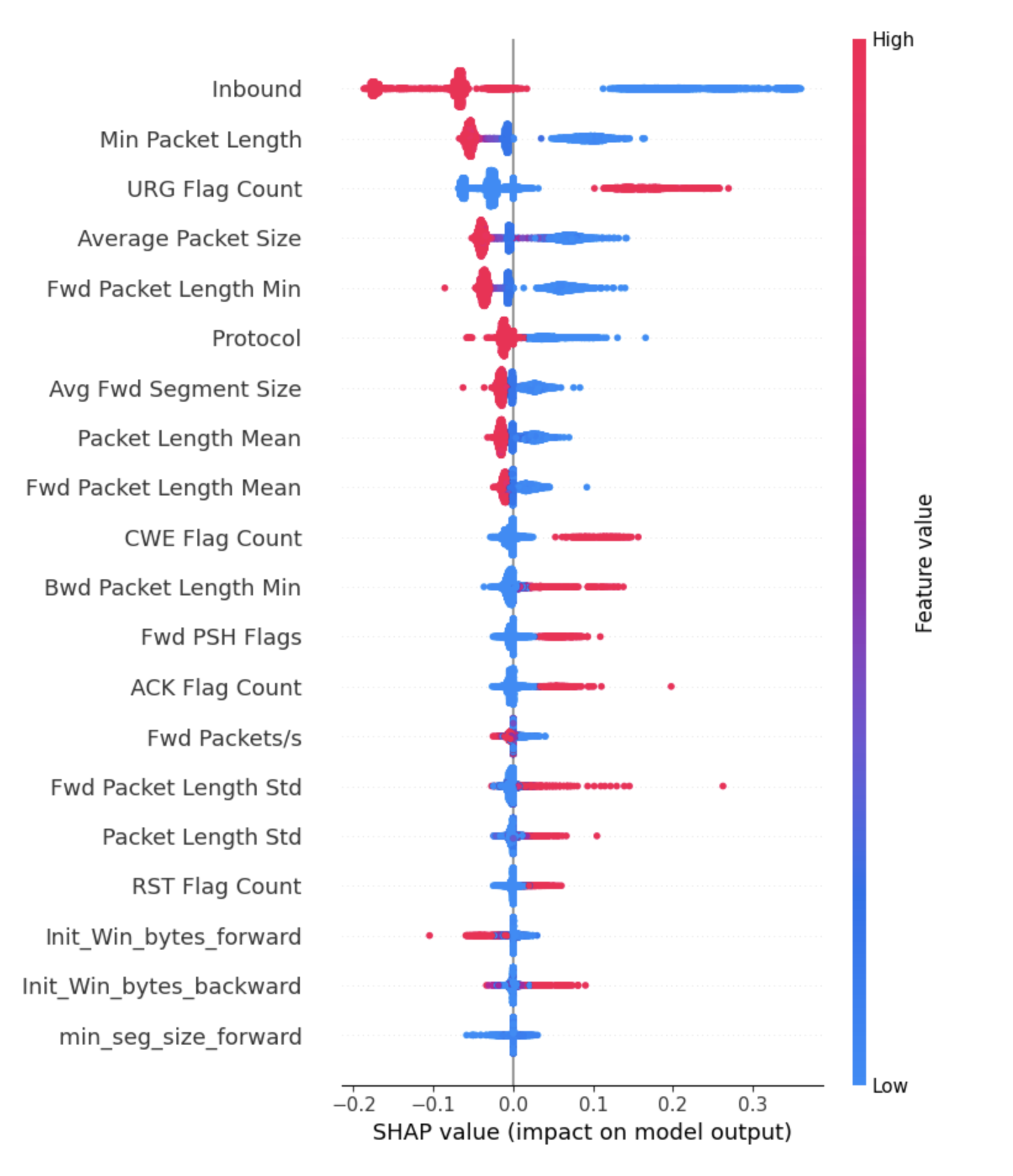} }}%% \par
    \par
    \subfloat[\label{fig: MFS-mean}\centering Mean Plot with Selected Features]{{\includegraphics[width=7cm]{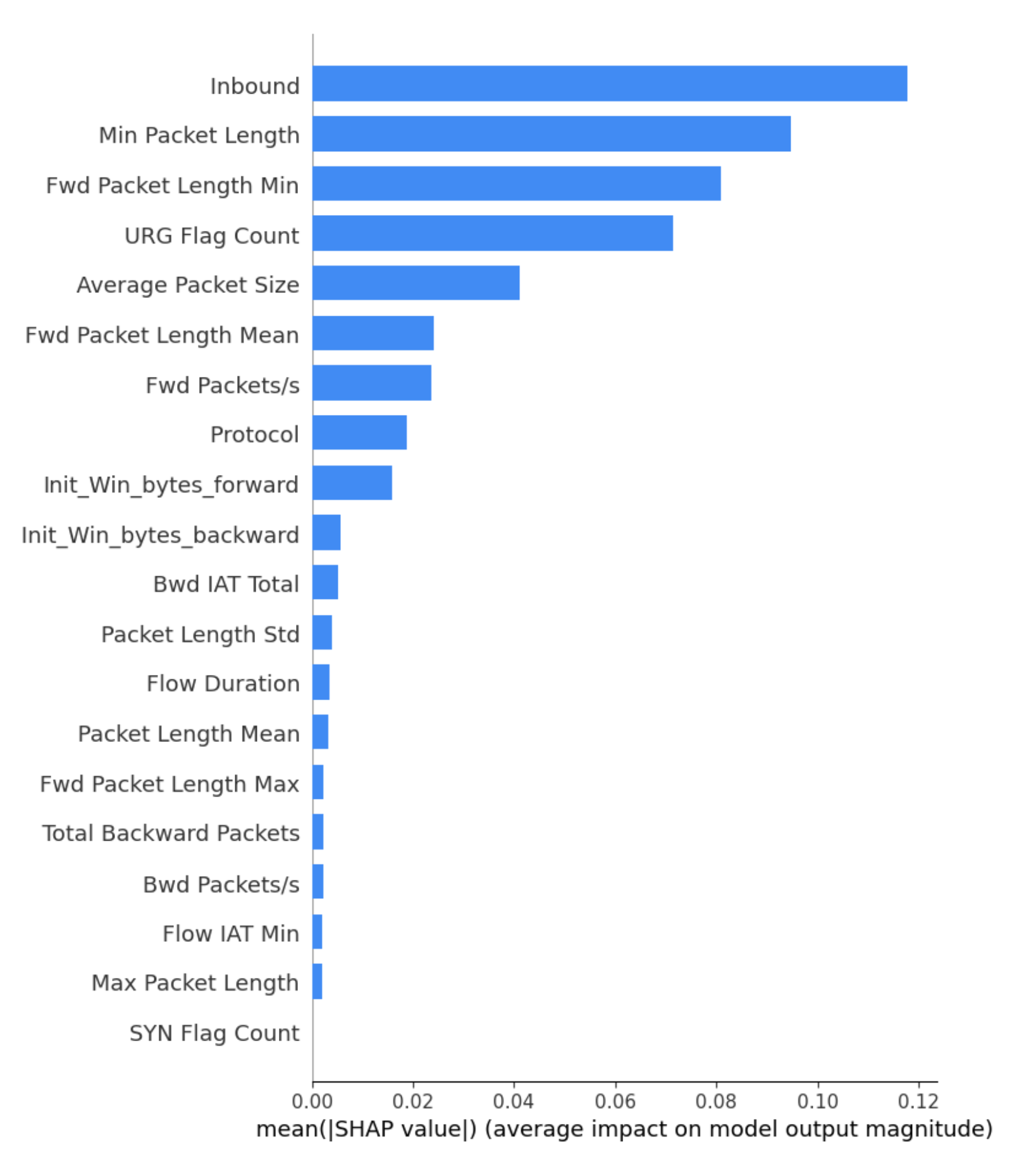} }}%    
    \subfloat[\label{fig: MFS-summary}\centering Summary Plot with Selected Features]{{\includegraphics[width=7cm]{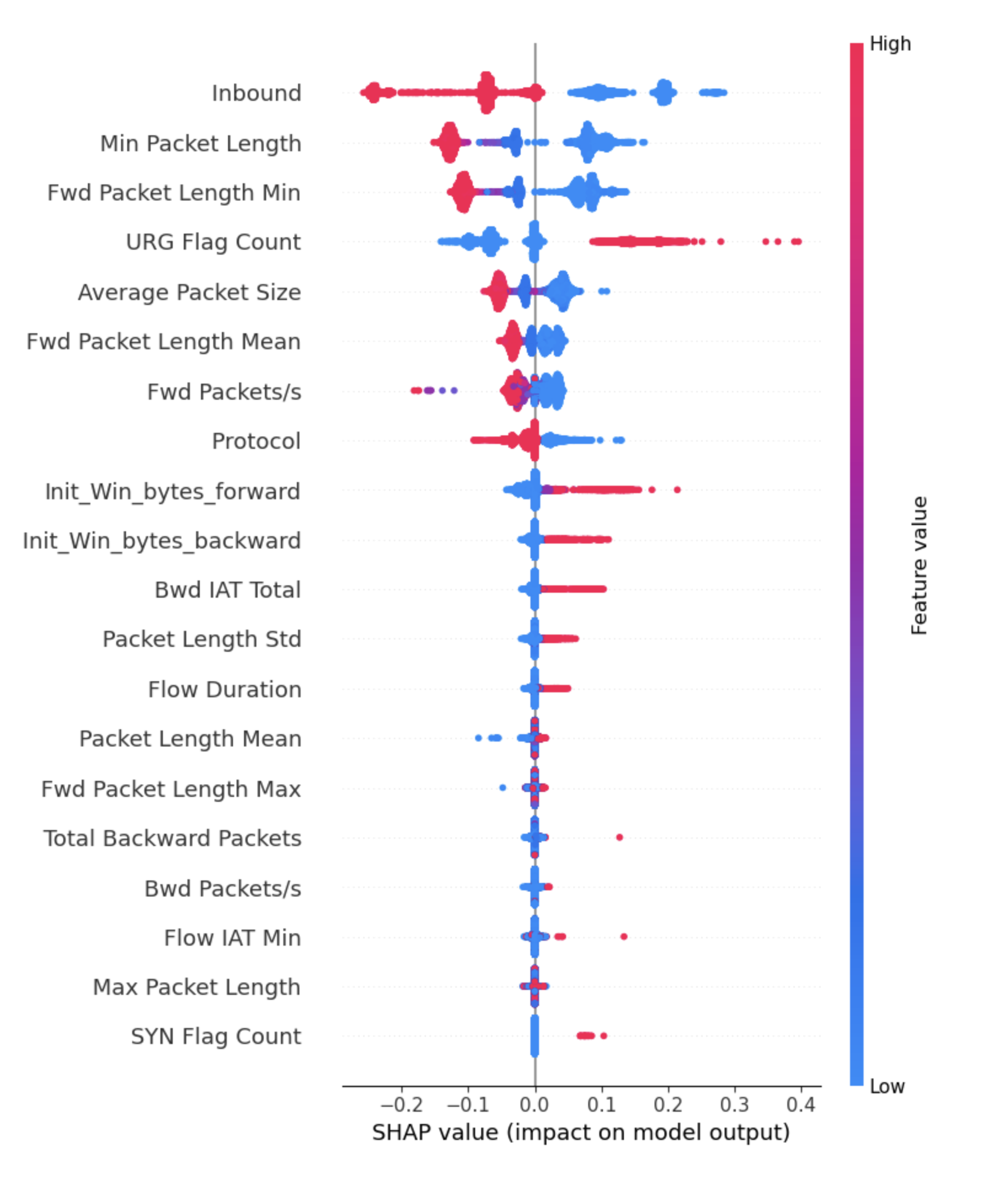} }}%
    \caption{One-to-All (Benign vs. four malicious types ): the Explanation of SHAP Values}%
    \label{fig: glo-All}%
\end{figure*}

Global explanation aims to be explained the feature contribution of the prediction results. In this study, we use the test set, which consists of a subset of approximately 0.001\% of the DDoS dataset, to generate explanations in the form of Shapley values. The global explanation of the prediction outcomes contains two scenarios between one-to-one and one-to-all explanations accompanied by all features and selected features separately. Note that the top 20 important features shown in the figure are plotted using all features. Thus, we select the top 20 important features for feature selection based on the proposed feature selection method (Fig.~\ref{fig: glo-One} and~\ref{fig: glo-All}).

\begin{figure*}[t]%
    \centering
    \subfloat[\label{fig: Odep1}\centering All Features: Top 1 Important Feature]{{\includegraphics[width=8cm]{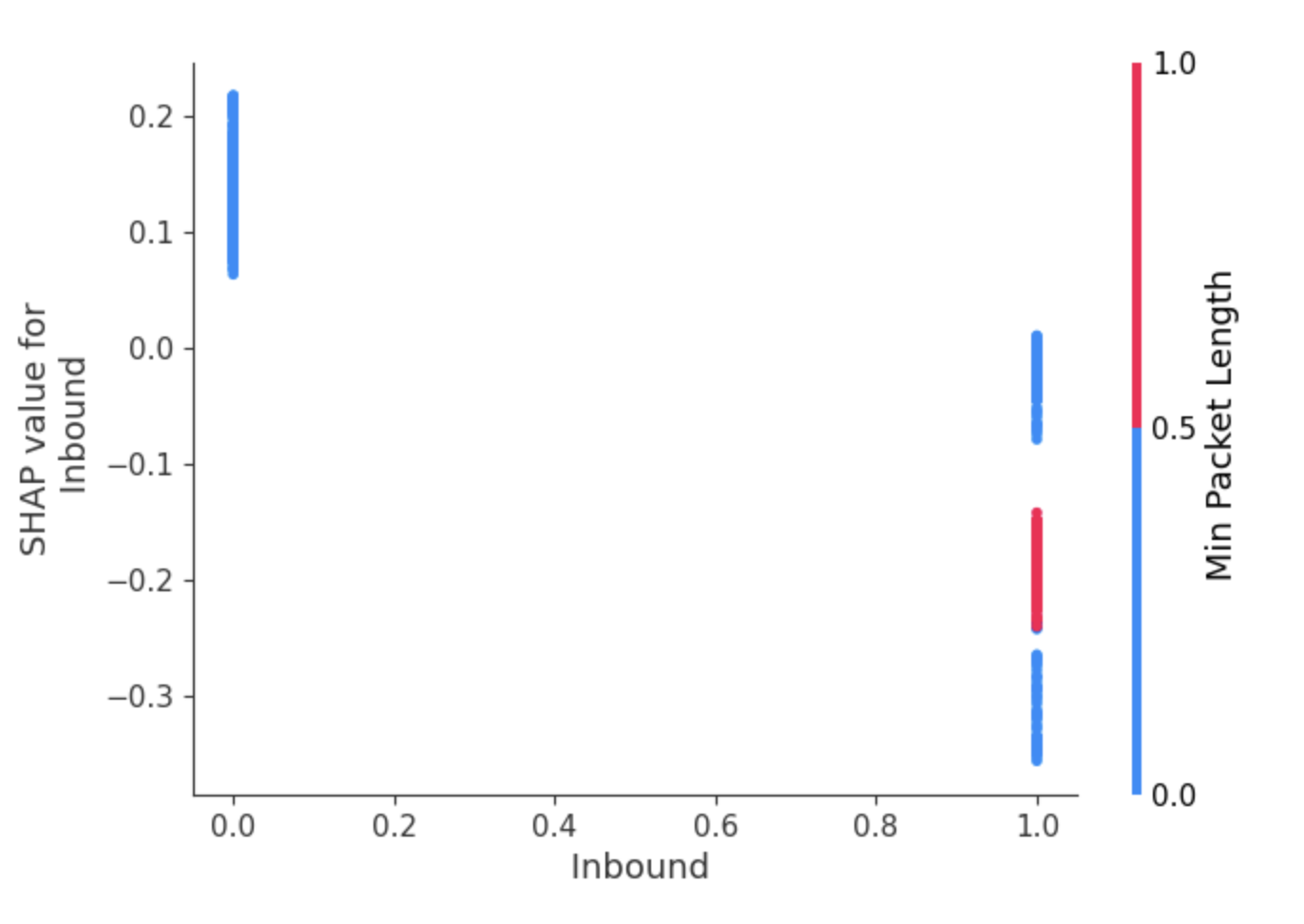} }}%
    \subfloat[\label{fig: Odep2}\centering Selected Features: Top 1 Important Feature]{{\includegraphics[width=8cm]{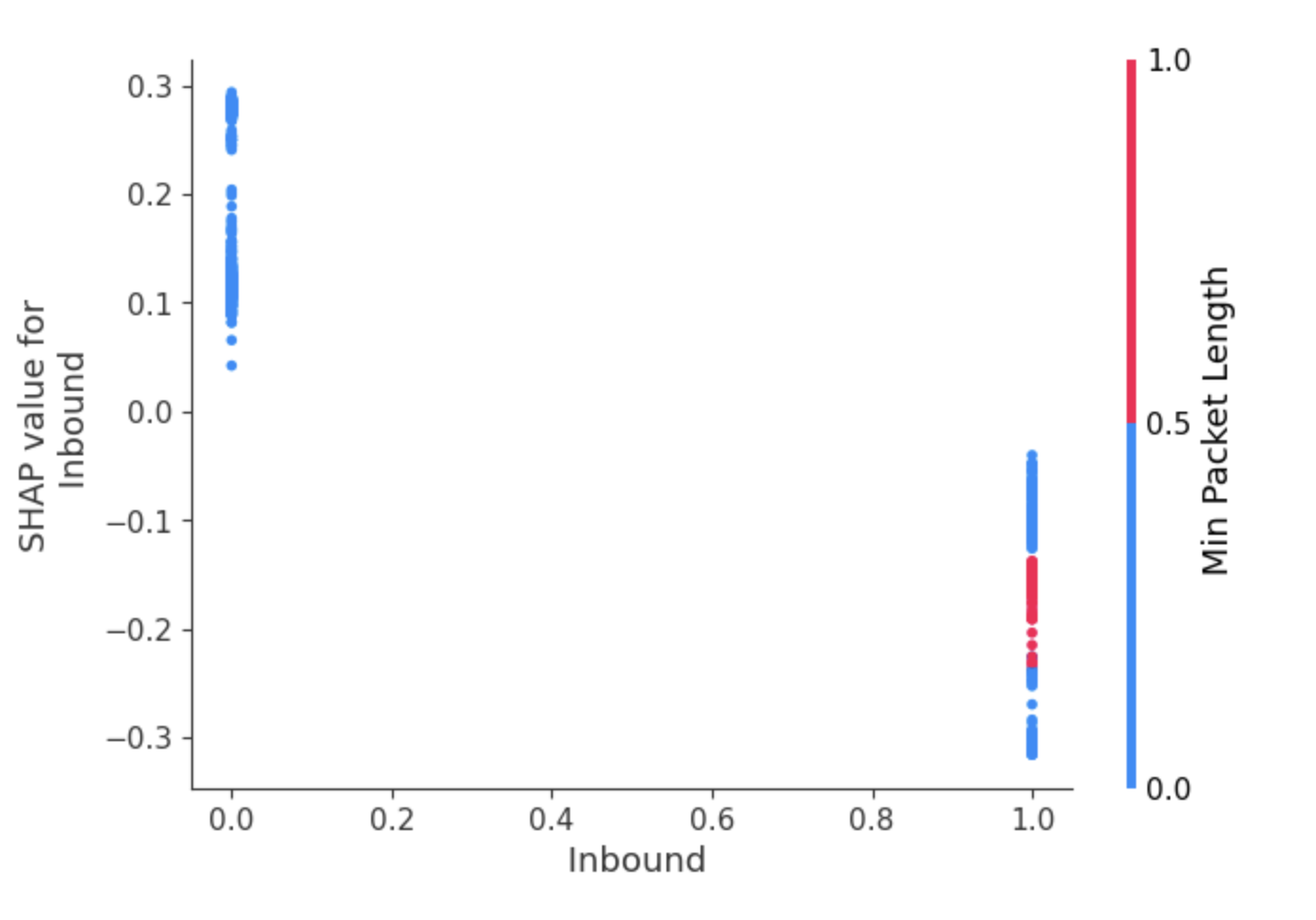} }}%
    \par
    \subfloat[\label{fig: Odep3}\centering All Features: Bottom 1 Feature]{{\includegraphics[width=8cm]{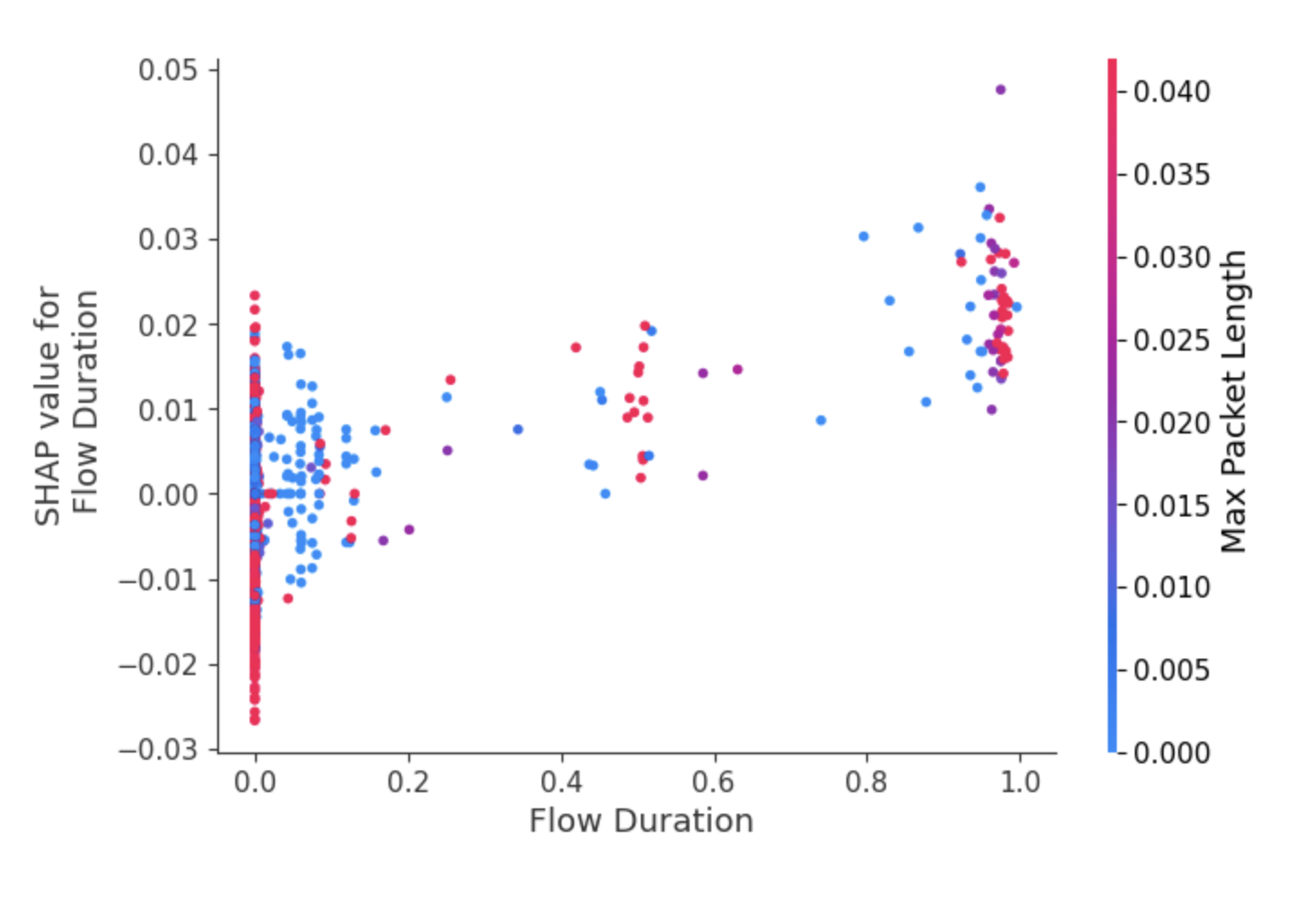} }}%
    \subfloat[\label{fig: Odep4}\centering Selected Features: Bottom 1 Feature]{{\includegraphics[width=8cm]{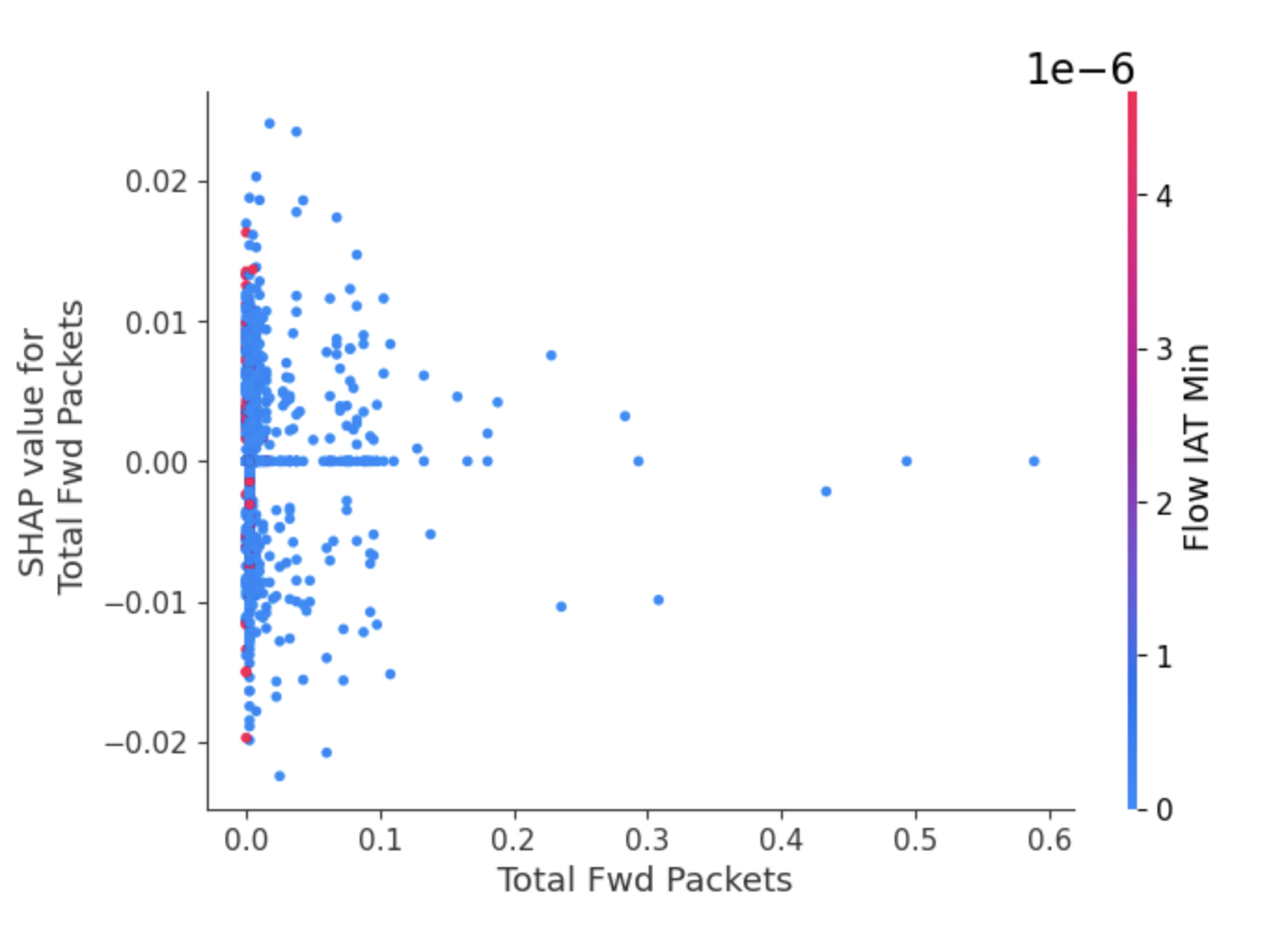} }}%
    \caption{Global Explanation: One-to-One Dependence Plot}%
    \label{fig: glo-Odependence}%
\end{figure*}

\begin{figure*}[t]%
    \centering
    \subfloat[\label{fig: Mdep1}\centering All Features: Top 1 Important Feature]{{\includegraphics[width=8cm]{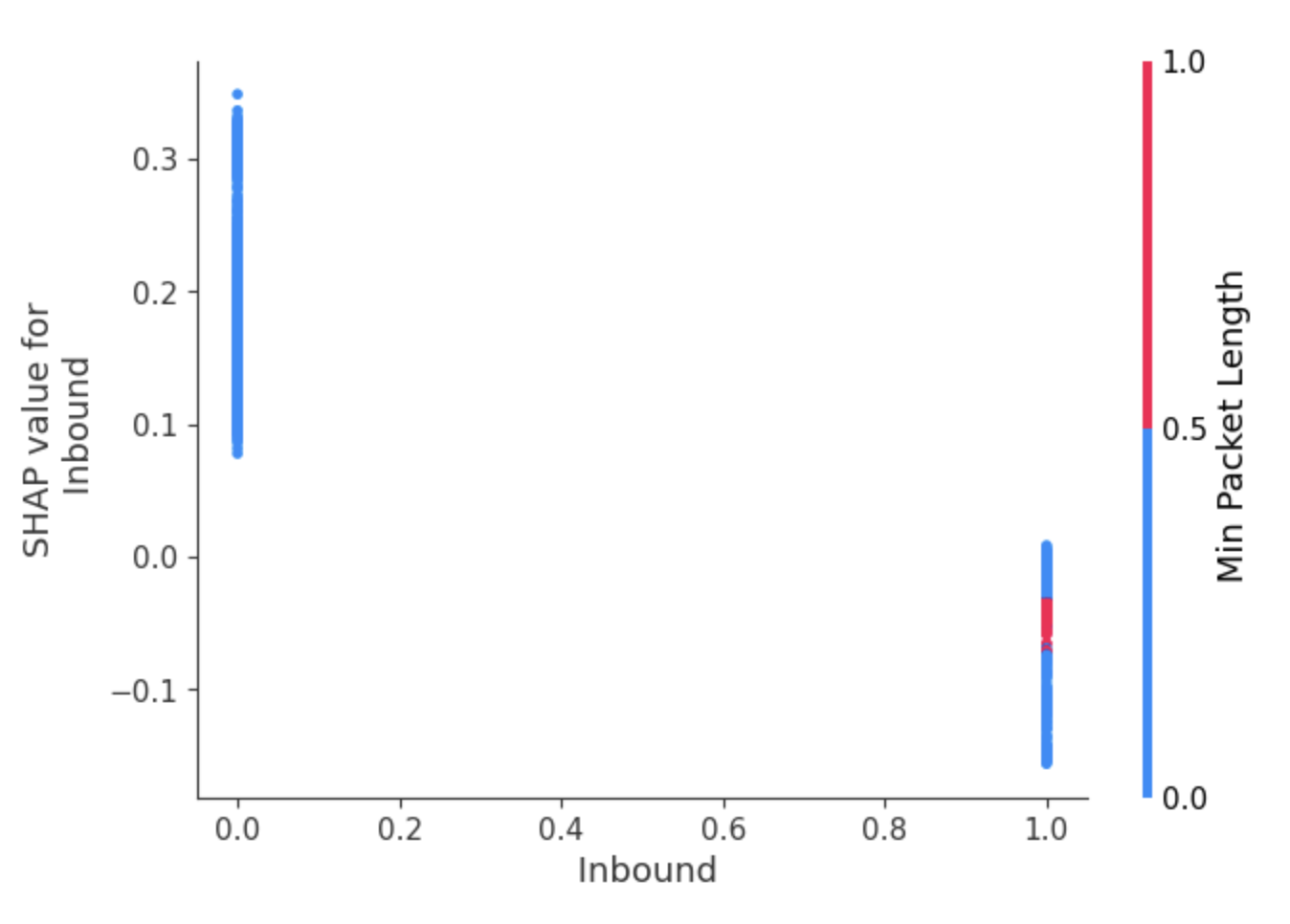} }}%
    \subfloat[\label{fig: Mdep2}\centering Selected Features:Top 1 Important Feature]{{\includegraphics[width=8cm]{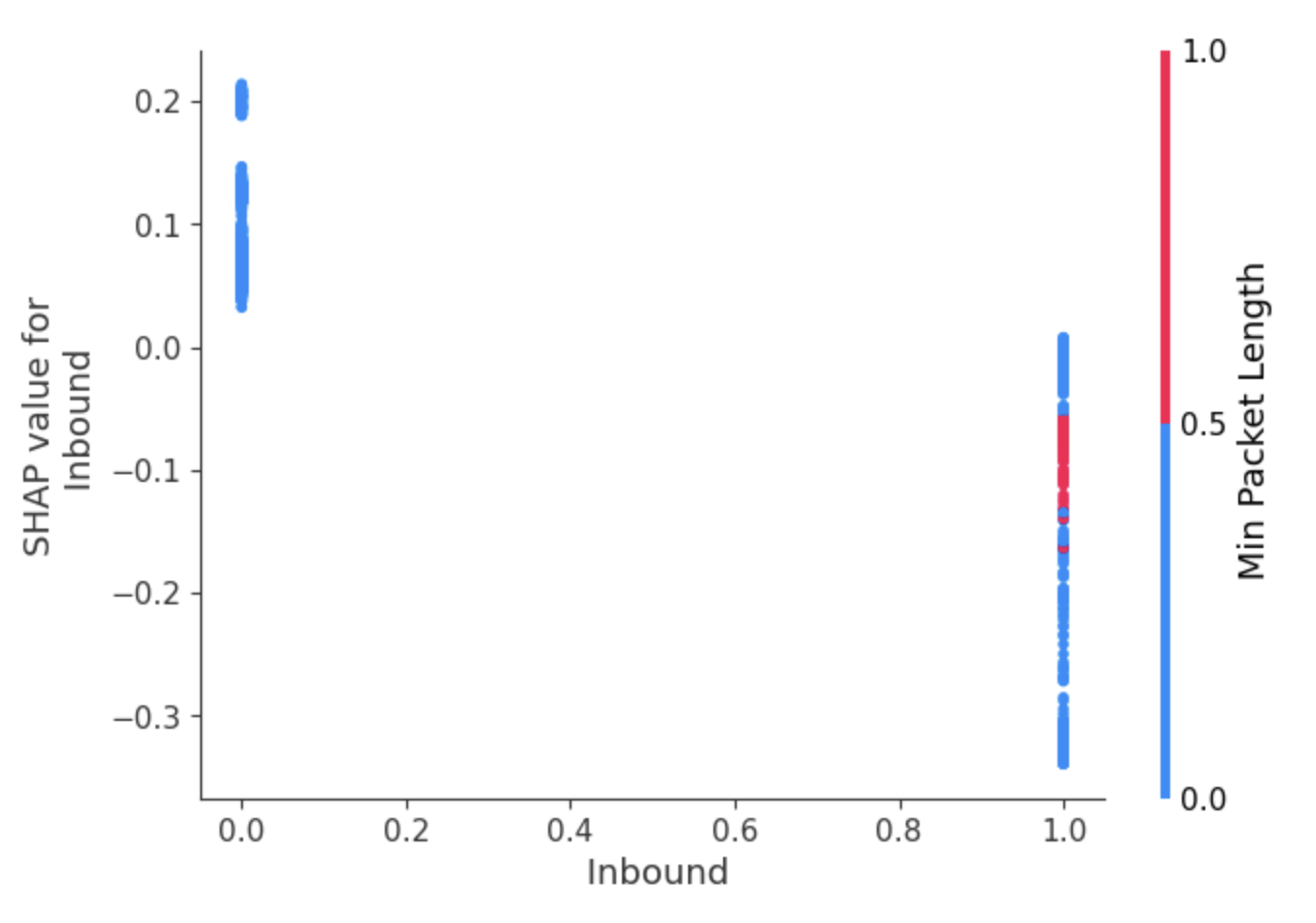} }}%
    \par
    \subfloat[\label{fig: Mdep3}\centering All Features: Bottom 1 Important Feature]{{\includegraphics[width=8cm]{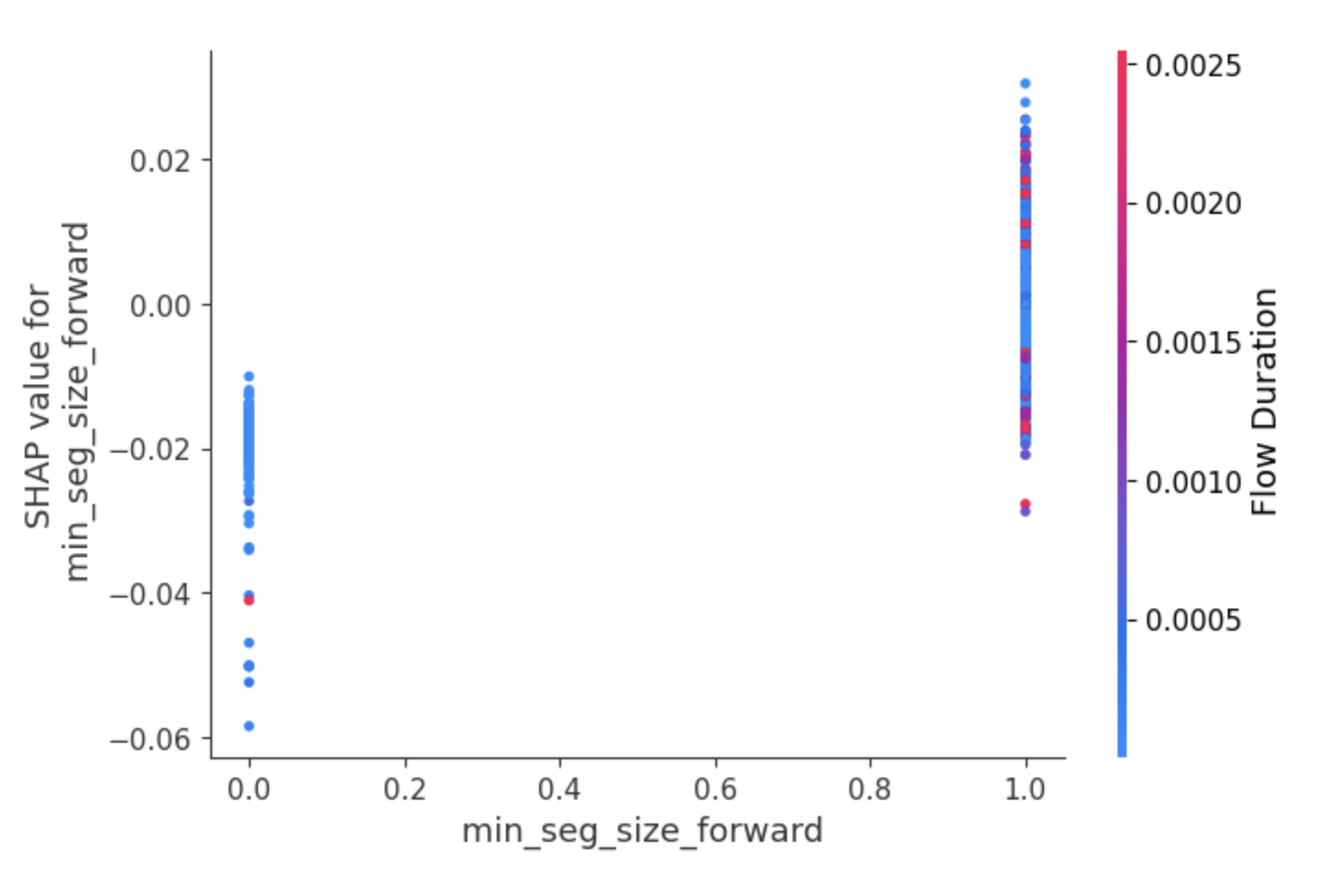} }}%
    \subfloat[\label{fig: Mdep4}\centering Selected Features:Bottom 1 Important Feature]{{\includegraphics[width=8cm]{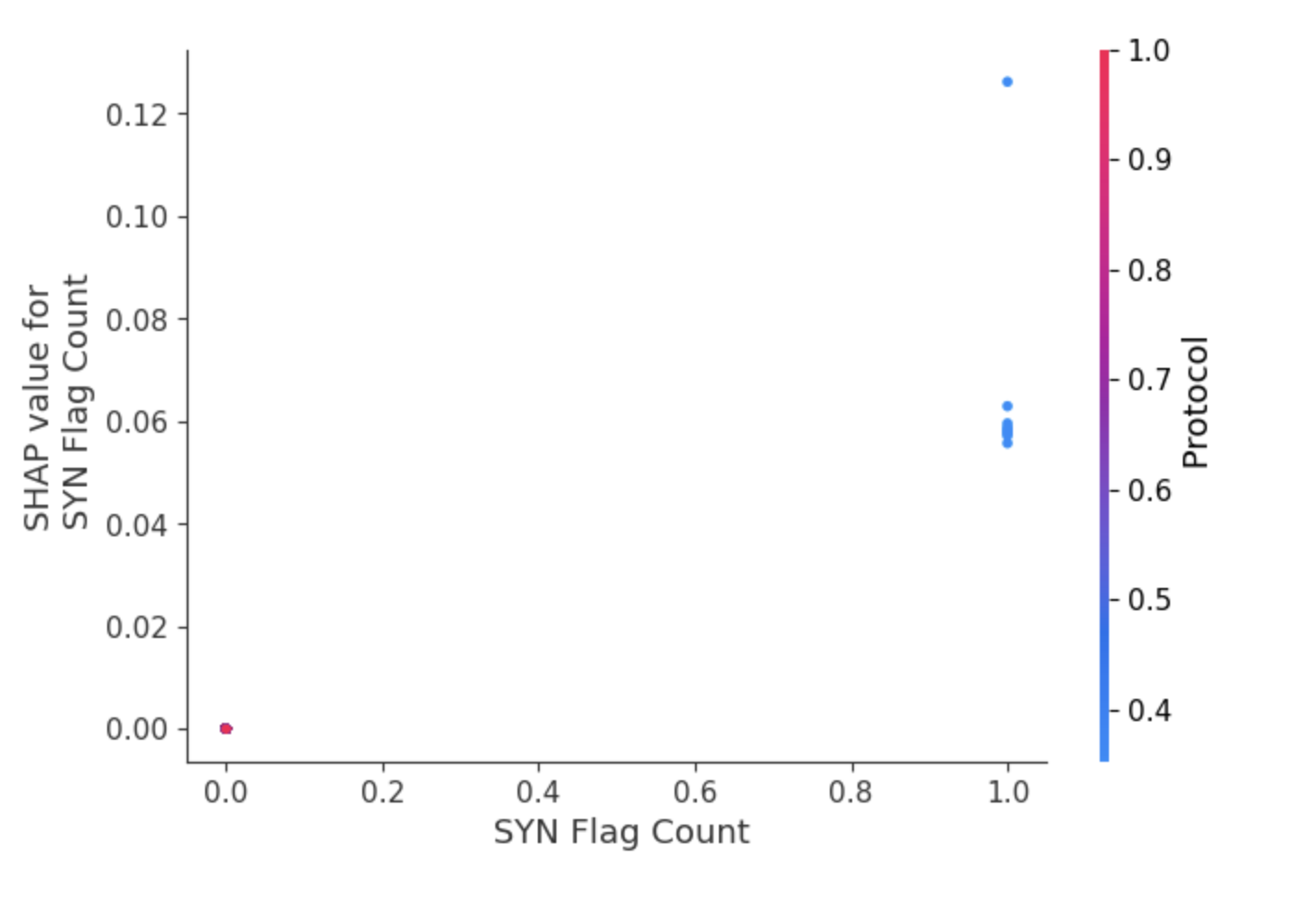} }}%
    \caption{Global Explanation: One-to-all Dependence Plot}%
    \label{fig: glo-Mdependence}%
\end{figure*}

Global SHAP explanation is commonly presented in two visualization plots. One of the most important plots in the explanation is the summary plot (Fig.~\ref{fig: glo-One} and~Fig.~\ref{fig: glo-All}). A dependence plot is another common global interpretation visualization plot (Fig.~\ref{fig: glo-Odependence} and~Fig.\ref{fig: glo-Mdependence}).

\textbf{Summary Plot}\par
Fig.~\ref{fig: OAll-mean} and \ref{fig: OFS-mean} show SHAP values in bar plots visualizing the one-to-one scenario based on the global importance of each feature with or without feature selection, where the global importance of each feature takes the mean absolute value for that feature across all given samples. Fig.~\ref{fig: OAll-summary} shows the visualization of the summary plot of the SHAP values for the top 20 contributing features by using all features from the testing DDoS dataset, while Fig.~\ref{fig: OFS-summary} depicts the summer plot by using the top 20 selected features. The large contribution of the top 20 features is shown on the left in descending order from high to low.

In the Fig.~\ref{fig: OAll-summary} and~Fig.~\ref{fig: OFS-summary} of the summary plots, a single point is plotted for each traffic flow (a dot represents a traffic flow). The SHAP values that fall on the left side of the x-axis have a negative impact on the prediction results, lowering the predicted values down and increasing the chances of being benign traffic. The values shown on the right side of the x-axis have a positive impact, increasing the predicted values and bringing them closer to the malicious traffic. In addition, the left vertical axis represents the feature names, ranked in descending order of all feature importance, while the right vertical axis represents the original value as it appears in the dataset and is colored in dots to represent high (red) or low (blue). Note that the horizontal values represent the SHAP values of the predictions that are associated with high or low predictions, while the vertical axis above point zero (0.0) represents no impact on the prediction results. For example, a SHAP value of zero (0.0) represents no impact on the prediction results, or close to zero represents low-quality predictions, while high-quality predictions, where SHAP values are far away from zero, also indicate either positive or negative correlations.

% For binary classification, red points indicate 1, while blue points indicate 0. 

\textbf{One-to-one classification explanation:}
Fig.~\ref{fig: OAll-mean} and \ref{fig: OFS-mean} show the mean SHAP value of all traffic flows, which represents the importance of features in decreasing order from high to low. The largest mean SHAP value had the highest importance among the others, which is a significant predictive feature for the prediction results of the proposed model. For example, the \textit{Inbount} feature is the most globally important feature in classifying DDoS traffic when using all features or the top 20 selected features. Although the \textit{URG Flag Count} is the third most important feature in the classification of the prediction results when using the top 20 selected features (Fig.~\ref{fig: OFS-mean}), it is the fifth most important feature in classifying benign and malicious traffic when using all features (Fig.~\ref{fig: OAll-mean}). However,  it is less important with the SHAP value of 0.05 in Fig.\ref{fig: OAll-mean} than the SHAP value of 0.06 in Fig.~\ref{fig: OFS-mean}. In addition, the feature  \textit{Flow Duration} is the least important feature in Fig.~\ref{fig: OAll-mean}. However, in Fig.\ref{fig: OFS-mean},~\textit{Flow Duration} is more important than among other features that fall under the~\textit{Flow Duration}. Finally, In Fig.~\ref{fig: OAll-mean} and Fig.~\ref{fig: OFS-mean}, the value of the feature~\textit{Inbound} is more important than about twice of the feature~\textit{Min Packet Length} on the prediction result. Overall, the importance of each feature has a smooth decreasing order in Fig.~\ref{fig: OAll-mean}, while the influence of feature importance decreases sharply from the sixth feature in Fig.~\ref{fig: OFS-mean}.

The \textit{Inbound} feature has the largest impact on both using all features and the selected top 20 feature classifications in Fig.~\ref{fig: OAll-summary} and~\ref{fig: OFS-summary}. Higher values (red) have a negative impact on the prediction results, pushing them toward benign, while lower values (blue) have a positive impact on the prediction results, pushing them toward malicious. The \textit{Min Packet Length}, \textit{Average Packet Size}, \textit{Fwd Packets/s} and \textit{Fwd Packet Length Min} features have a similar impact tendency as \textit{Inbound}, but these features have a different contribution impact order in Fig.~\ref{fig: OAll-summary} and~\ref{fig: OFS-summary} with or without feature selection. 

The higher SHAP values of the \textit{URG Flag Count} feature have a higher positive impact on the prediction results, pushing it towards malicious, while the lower SHAP values have a negative impact, pushing them towards benign, as shown in both Fig.~\ref{fig: OAll-summary} and Fig.~\ref{fig: OFS-summary}. Furthermore, in the feature selection scenario, the \textit{URG Flag Count} feature has a smaller impact than the use of all features.

In Fig.~\ref{fig: OFS-summary}, the smallest contribution of \textit{Flow Duration}  has the least impact on the classification prediction results because of the SHAP values being closer to zero and no significant effect. However, the \textit{Total Fwd Packets} feature has no impact or less impact than most of the top 20 features in Fig.~\ref{fig: OAll-summary} because this feature is outside the range of the largest contribution of the top 20 features.

\textbf{One-to-all classification explanation:}
Fig.~\ref{fig: glo-All} shows the mean SHAP value plot and the summary plot of the traffic flows in the one-to-all scenario, which also represents the feature importance and the most contributing features separately in decreasing order from high to low. 

Fig.~\ref{fig: MAll-mean} and~Fig.~\ref{fig: MFS-mean} show the global feature importance in bar plots, representing the importance of the features in decreasing order from high to low by using all features and selected features respectively. A large mean SHAP value on the x-axis illustrates the most predictive features for the classification. The features \textit{Inbound}, \textit{Min Packet Length} are the top 2 important predictive features for the classification predictions in both Fig.~\ref{fig: MAll-mean} and and~Fig.~\ref{fig: MFS-mean}. The features \textit{Inbound} is the largest important feature, and the mean SHAP value is greater than 0.12. However, the importance of the feature \textit{Min Packet Length} with all features plotted is not as significant importance as with the selected features. This is because the mean SHAP value in the plot of all features is 0.07, compared to 0.09 in the plot of the selected feature. Similarly, the importance of each feature has a smooth decreasing order from the feature \textit{Min Packet Length} in Fig.~\ref{fig: MAll-mean}, while the influence of feature importance decreases sharply from the fifth feature in Fig.~\ref{fig: MFS-mean}. The least important feature is \textit{Min\_seg\_size\_forward} in Fig.~\ref{fig: MAll-mean}, while the \textit{SYN Flag Count} feature is the least important feature in Fig.~\ref{fig: MFS-mean}.

The summary plot of the one-to-all scenario shows in ~Fig.~\ref{fig: MAll-summary} and~\ref{fig: MFS-summary}. The top 5 most contributing features (except feature \textit{URG Flag Count}) in ~Fig.~\ref{fig: MAll-summary} and~\ref{fig: MFS-summary} have a greater impact on the prediction results. Higher values have a negative impact on the prediction results, pushing them toward benign, while lower values have a positive impact on the prediction results, pushing them toward malicious. In contrast, the feature with a higher impact on the prediction result is \textit{URG Flag Count}. The higher values have a positive impact, pushing it toward benign, while lower values have a negative impact, pushing it toward malicious. 

The higher value of the \textit{Packet Length Mean} feature has a negative impact in Fig.~\ref{fig: MAll-summary}, whereas it has a small positive impact in Fig.~\ref{fig: MFS-summary}. The higher values of the \textit{Init\_win\_bytes\_forward} and \textit{Init\_win\_bytes\_backward} have a higher positive impact on the prediction results in both Fig.~\ref{fig: MAll-summary} and Fig.~\ref{fig: MFS-summary}, pushing them towards to benign on the model decision. However, those two features contribute more impact on the prediction results in selected features shown in Fig.\ref{fig: MFS-summary}, compared to the contribution in Fig.\ref{fig: MAll-summary}.

\textbf{Dependence plot}\par
The dependence plot is an important plot that can provide more information about the relationship between the feature values and the corresponding SHAP values for each traffic flow. A SHAP dependence plot is a scatter plot that illustrates the impact of a single feature on the predictions made by the proposed model. Each point represents a single prediction from the entire dataset. The horizontal axis represents the value of the features, while the left vertical axis illustrates the SHAP value of the corresponding feature, representing how much of the features impact the prediction results. The right vertical axis represents the effect of the interaction with the horizontal plot feature, highlighted in red color (Note that this interaction feature is chosen automatically).

Fig.~\ref{fig: glo-Odependence} shows the most and least important features before and after feature selection in a one-to-one scenario. The SHAP dependence plot with and without feature selection for the top 1 importance feature of \textit{Inbound} are shown in Fig.~\ref{fig: Odep1} and~\ref{fig: Odep2}. In these two figures, the negative value of SHAP values (e.g. -0.1) with the \textit{Inbound} value (0) has a great impact on the malicious. Meanwhile, there are a large number of positive SHAP values (e.g. 0.1) localized with the \textit{Inbound}feature value (1), as shown in Fig.~\ref{fig: Odep1} and~\ref{fig: Odep2}, which has the large impact of the benign traffic. In contrast to the top one feature impact of the prediction results, Fig.~\ref{fig: Odep3} and~~\ref{fig: Odep4} show the lowest botten (bottom 1) influence of the prediction results. The dots are distributed around the SHAP values of 0.0, which has a small influence on the prediction results.

In a one-to-all scenario, the dependence plots of the top 1 important feature with and without feature selection are shown in Fig.~\ref{fig: Mdep1} and~\ref{fig: Mdep2} separately. The most contributing feature \textit{Inbond} has both negative and positive SHAP values, corresponding to \textit{Inbond} values of 1.0 and 0.0, which have influenced and pushed towards the benign and malicious traffic respectively. Similarly, the interaction feature \textit{Min Packet Length} takes the value 1 attached as the \textit{Inbond} value of 1.0, which is likely to be benign traffic. In contrast to the top 1 contributing feature, the bottom one impacted feature, as shown in Fig.~\ref{fig: Mdep3} and~~\ref{fig: Mdep4}. There is a large number of dots distributed closer to the SHAP value of 0.0, which has a smaller influence on the prediction results.

% \begin{figure}[h]%
%     \centering
%     \label{fig:summary-multiple}
%     \subfloat[\centering Summary Plot]{{\includegraphics[width=9cm]{Figures/one-multiple.png} }}%
%     % \qquad
%     \par
%     \label{fig:mean-multiple}
%     \subfloat[\centering Mean plot]{{\includegraphics[width=9cm]{Figures/Global-multiple.png} }}%
%     \caption{One-to-Multiple: The Explanation of SHAP Values Plot of MLP Classifier Prediction}%
%     \label{fig:glo-multiple}%
% \end{figure}

\begin{figure*}[t]%
    \centering
    \subfloat[\label{fig: OAll1}\centering Malicious (DNS): Predicted Malicious (DNS)]{{\includegraphics[width=9cm]{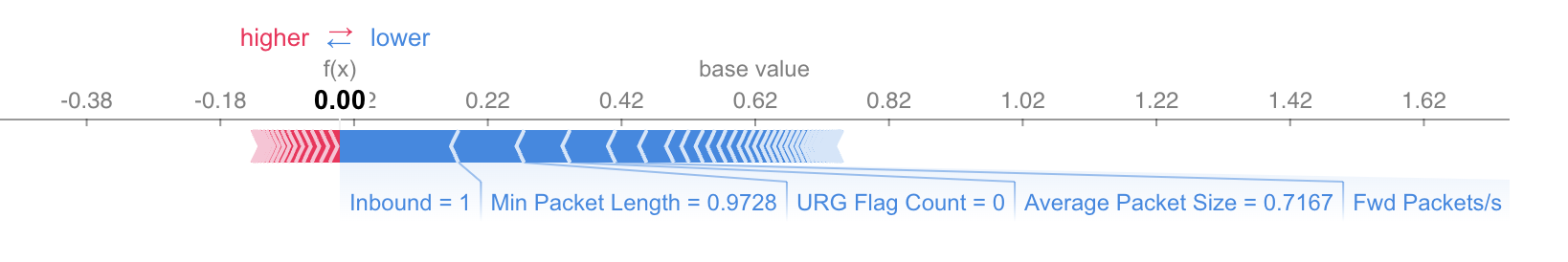} }}%
    \subfloat[\label{fig: OAll2}\centering Malicious (DNS): Predicted Benign]{{\includegraphics[width=9cm]{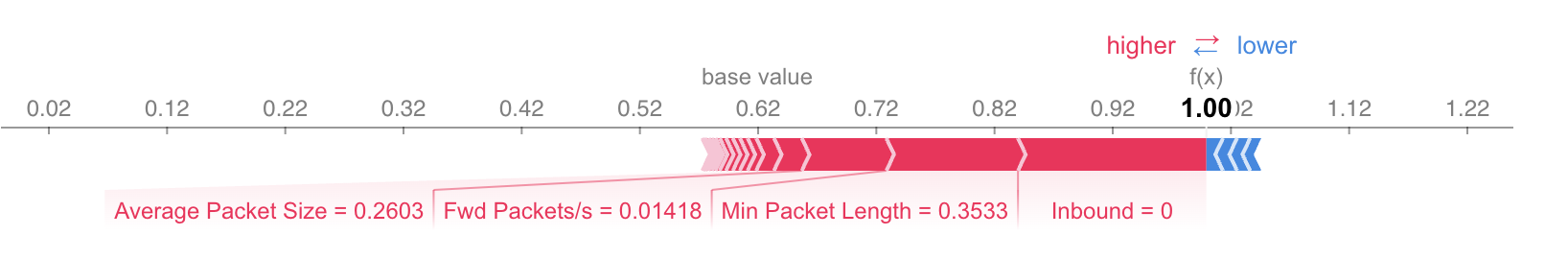} }}%
    \par
    \subfloat[\label{fig: OAll3}\centering Benign: Predicted Benign]{{\includegraphics[width=9cm]{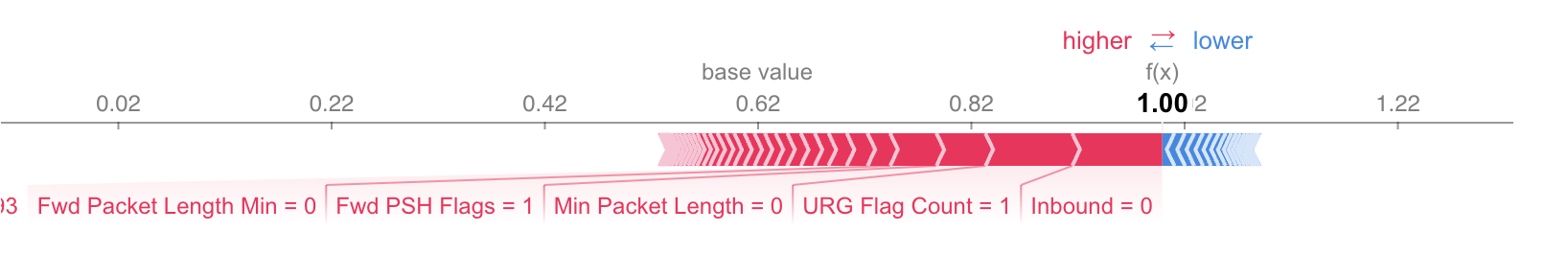} }}%\par
    \subfloat[\label{fig: OAll4}\centering Benign: Predicted Malicious (DNS)]{{\includegraphics[width=9cm]{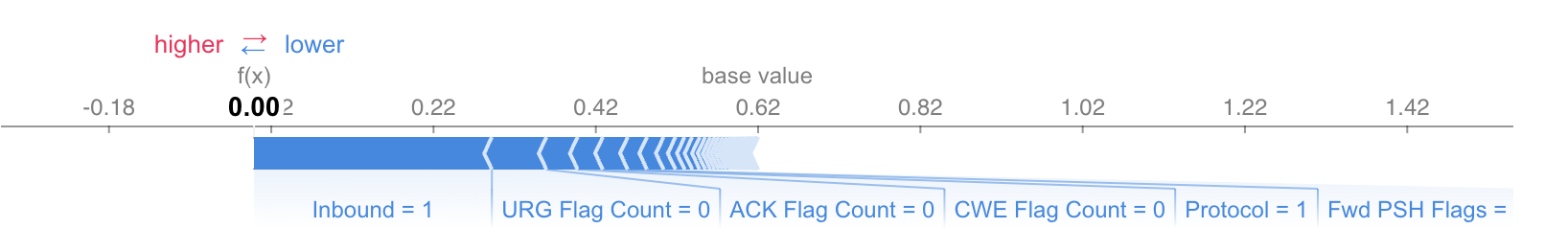} }}%
    \caption{Local Explanation: One-to-one with All Features}%
    \label{fig: Loc-OAll}%
\end{figure*}

\subsubsection{Local Explanation}
The local explanation can explain the local interpretability, provide an explanation for each individual sample and point out the contribution of features in explaining how a model makes decisions - based on Shapley values computed by SHAP, which can make the proposed model more convincing for the prediction results. Each feature value represents a force that either increases or decreases the prediction results.

\textbf{Force Plot}\par
One of the most important plots in the local explanation is the force plot. The force plot allows us to visualize the contribution of the features to the classification prediction results for a particular traffic flow. In the force plots, the base value is the average of all the prediction result values, starting from the baseline (the boundary between red and blue). The red color represents features that push the prediction results toward higher, while the blue color represents features that push the prediction results toward lower. The length of the feature under the horizontal line determines the magnitude of the impact (e.g. the larger arrow block of the color, the greater impact of the feature on the prediction results). 

In this study, we use the SHAP explanation for interpreting the prediction of a specific traffic flow. The explanation can be divided into two scenarios to explain.
% in the following subsection.

% The visualization of the "force" is the attribution feature of the Shapley values. The Shapley value is represented separately by red and blue arrows.

\begin{figure}[h]%
    \centering
    \subfloat[\label{fig: OFS1}\centering Malicious (DNS): Predicted Malicious (DNS)]{{\includegraphics[width=9cm]{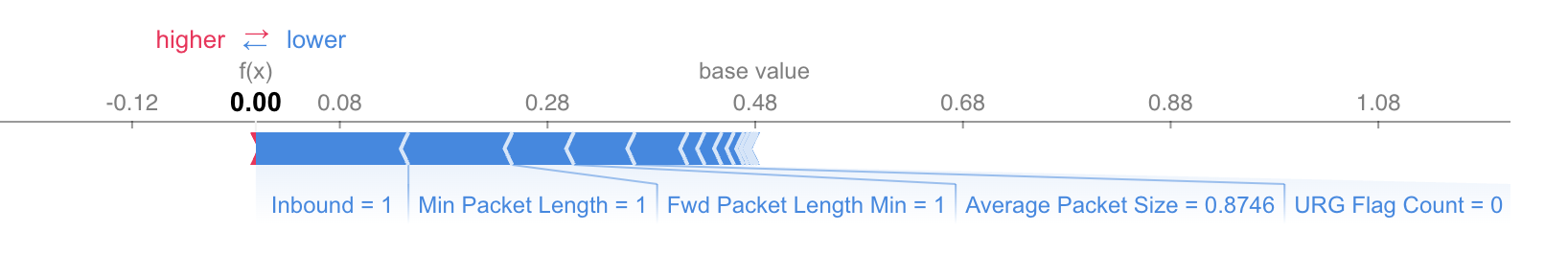} }}%
    \par
    \subfloat[\label{fig: OFS2}\centering Benign: Predicted Benign]{{\includegraphics[width=9cm]{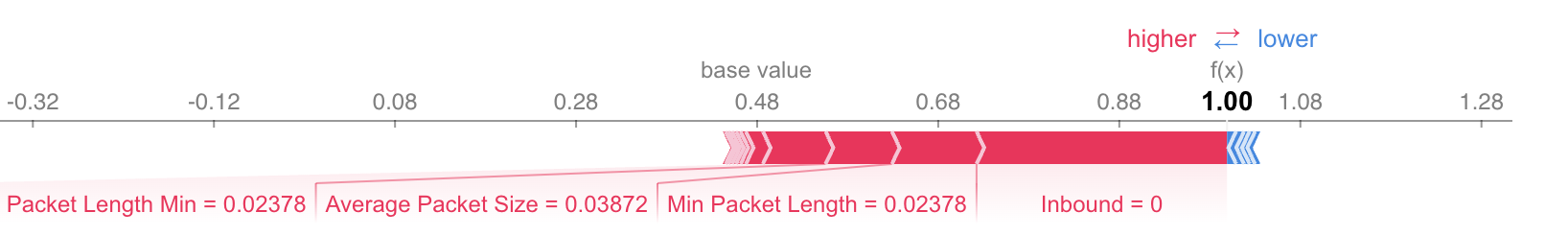} }}%
    \par
    \subfloat[\label{fig: OFS3}\centering Benign: Predicted Malicious (DNS)]{{\includegraphics[width=9cm]{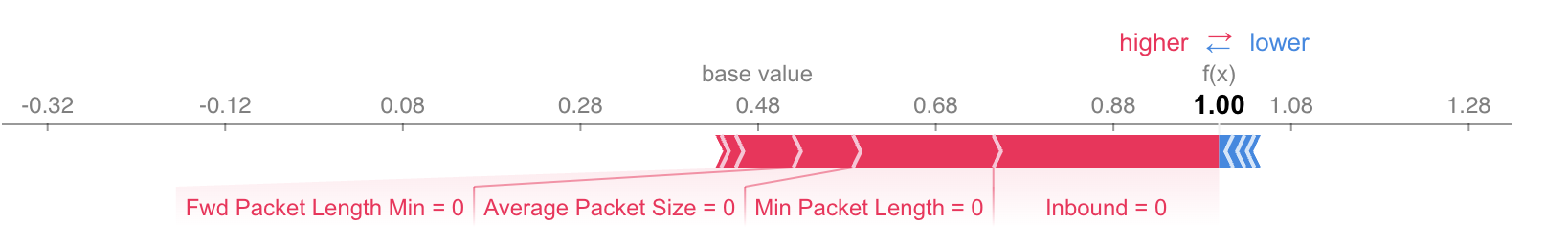} }}%
    % \label{fig:mean-one}
    % \par
    % \subfloat[\centering All Features with FP]{{\includegraphics[width=18cm]{Figures/OAll-10.png} }}%
    \caption{Local Explanation: One-to-one with Feature Selection}%
    \label{fig:Loc-OFS}%
\end{figure}

\begin{figure*}[t]%
    \centering
    \subfloat[\label{fig: MAll1}\centering Malicious: Predicted Malicious]{{\includegraphics[width=9cm]{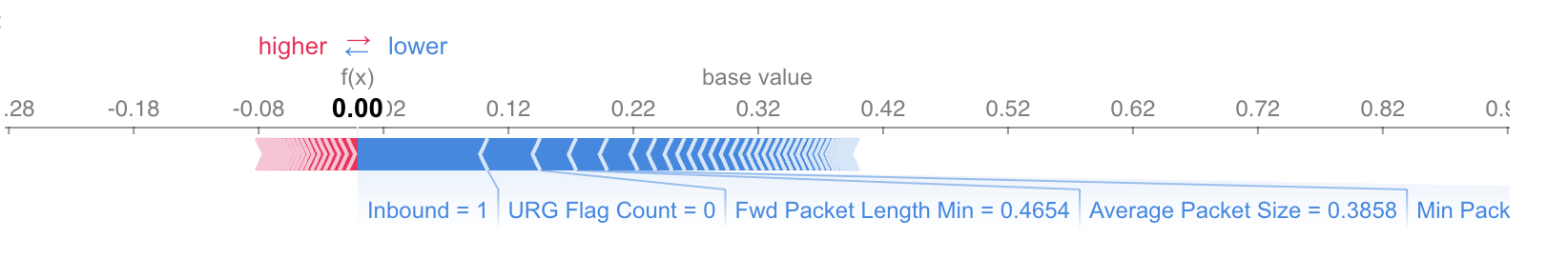} }}%
    \subfloat[\label{fig: MAll2}\centering Malicious: Predicted Benign]{{\includegraphics[width=9cm]{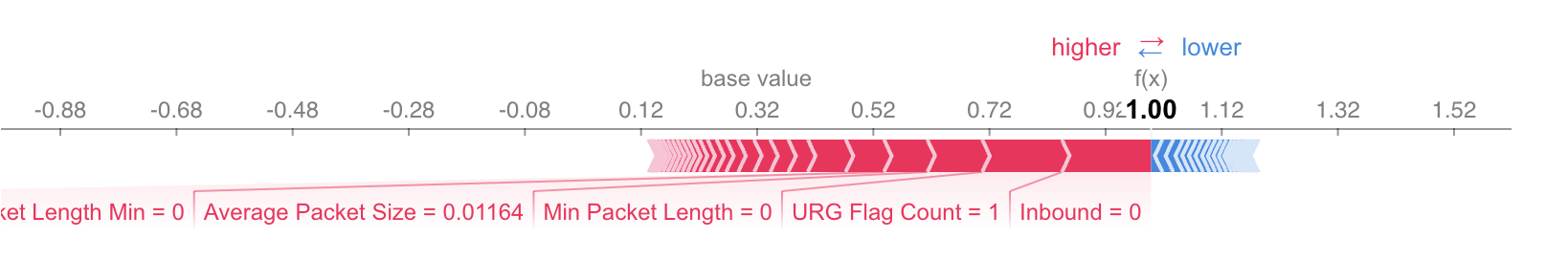} }}%
    \par
    \subfloat[\label{fig: MAll3}\centering Benign: Predicted Benign]{{\includegraphics[width=9cm]{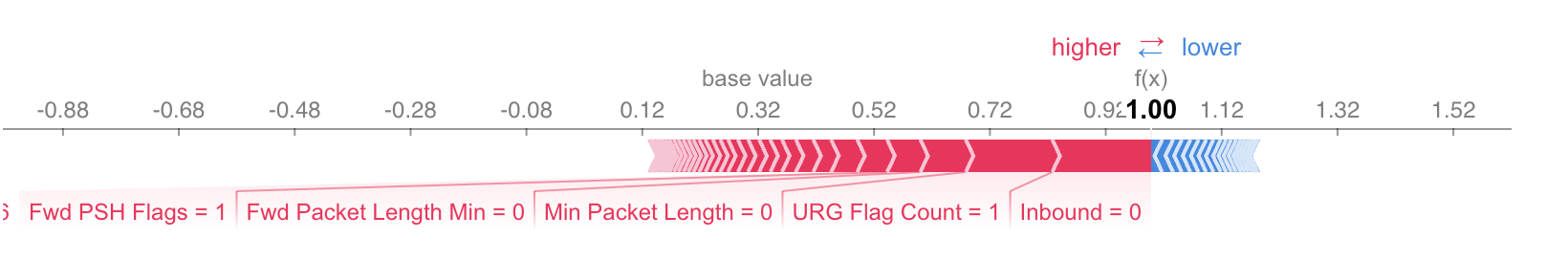} }}%
    \subfloat[\label{fig: MAll4}\centering Benign: Predicted Malicious]{{\includegraphics[width=9cm]{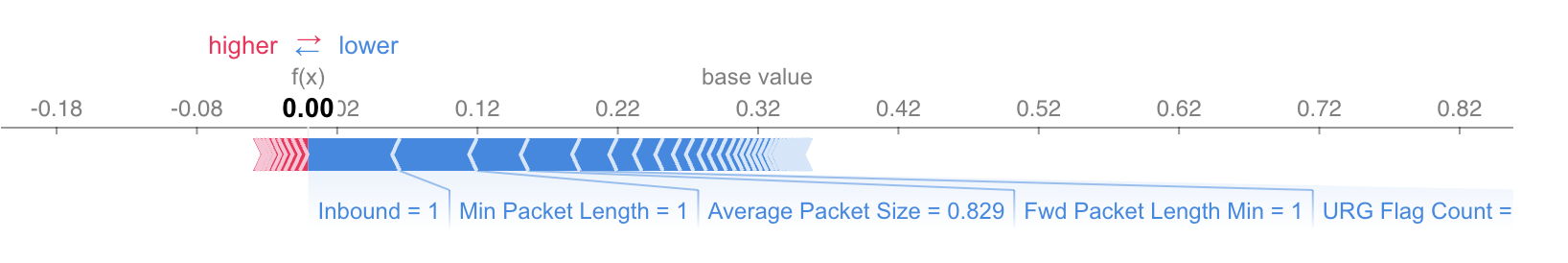} }}%
    \caption{Local Explanation: One-to-all with All Fearues}%
    \label{fig: Loc-MAll}%
\end{figure*}

\textbf{One-to-one scenario:}
Fig.~\ref{fig: Loc-OAll} shows the force plot on the one-to-one scenario using all the features from the test set of the DDoS dataset, which can explain the contribution of each feature to the prediction results. The base value here is 0.62. Fig.~\ref{fig: OAll1} shows a malicious sample, where the traffic is correctly predicted to be malicious, while Fig.~\ref{fig: OAll3} is a benign sample, which is correctly classified as benign. Furthermore, in Fig.~\ref{fig: OAll1}, the top 3 features (in order of its values) \textit{Inbound} (1), \textit{Min Packet Length} (0.9728), \textit{URG Flag Count} (0) have the most negative impact (blue) on the prediction results, pushing it towards to malicious (0), while the top 3 features \textit{Inbound} (0), \textit{URG Flag Count} (1), \textit{Min Packet Length} (0) have the most greater positive impact (red), pushing it towards benign (1). However, Fig.~\ref{fig: OAll2} and Fig.~\ref{fig: OAll4} show that the traffic flow cannot be classified correctly. For example, in Fig.~\ref{fig: OAll2}, the traffic flow is malicious, but it has been misclassified as benign, with the largest positive impact features (\textit{Inbound} (0), \textit{Min Packet Length} (0.3533), etc, pushing it towards benign (1). Similarly, a benign sample has been misclassified as malicious, because the features, including \textit{Inbound} (1), \textit{URG Flag Count} (0), etc have a negative impact and push it towards malicious(0). 

In the feature selection scenario, Fig.~\ref{fig: OFS1} and ~\ref{fig: OFS2} show that the traffic flows are correctly classified as malicious and benign traffics separately. In Fig.~\ref{fig: OFS1}, the most significant impact features are negative impacts (blue), including \textit{Inbound} (1), \textit{Min Packet Length} (1), etc., which increase the prediction result and push it towards malicious (0.00). Similarly, Fig.~\ref{fig: OFS2} shows the positive impact (red) of the features, including the features \textit{Inbound} with the value of 0, \textit{Min Packet Length} with the value of 0.02378, etc, which push it toward to benign (1.00). Furthermore, Fig.~\ref{fig: OFS3} is one the benign traffic misclassified as malicious with the positive impact features, including the feature \textit{Inbound} with the value of 0, \textit{Min Packet Length} with the value of 0, etc, which push it toward to benign (1.00). Note that all malicious traffic can be correctly classified as malicious.

\textbf{One-to-all scenario:}
Fig.~\ref{fig: Loc-MAll} shows force plots of a one-to-all scenario based on local explanation with all features, explaining benign and malicious traffic separately. 

Fig.~\ref{fig: MAll1} and~\ref{fig: MAll3} show that the traffic flows are correctly classified as malicious and benign separately using all features. The negative impact is shown in blue color in Fig.~\ref{fig: MAll1} which pushes the prediction result towards malicious (0.00). The feature that contributes the most to increasing the prediction value is \textit{Inbound} with a value of 1. The other higher contribution comes from the following features, including \textit{URG Flag Count} (0), \textit{Fwd Packet Length Min} (0.4654), etc. Similarly, the positive impact is shown in red color (Fig.~\ref{fig: MAll3}), which pushes the prediction towards benign (1.00) with the features \textit{Inbound} (0), \textit{URG Flag Count} (1) and \textit{Min Packet Length} (0), etc.  In contrast, Fig.~\ref{fig: MAll2} and~\ref{fig: MAll4} show that the traffic flows are misclassified. For example, the largest contribution of the \textit{Inbound} in Fig.~\ref{fig: MAll2} with a value of 0 can cause this malicious traffic to be misclassified as benign. Fig.~\ref{fig: MAll3} shows a similar impact tendency (the largest impact feature \textit{Inbound} with a value of 1) to cause the malicious traffic to be misclassified as benign.

Fig.~\ref{fig: OFS1} and ~\ref{fig: OFS2} show that in the feature selection scenario, the traffic flows are correctly classified as malicious and benign separately. The features \textit{Min Packet Length} (0.9837), \textit{Fwd Packet Length Min} (0.8448), and \textit{Inbound} (1), etc have a negative impact (blue), and push the prediction results toward malicious (0.00). Whereas, in ~\ref{fig: OFS2}, the features \textit{Inbound} (0), \textit{Fwd Packet Length Min} (0.02275), and \textit{Min Packet Length} (0.02649) have a greater positive impact (red) on the prediction results, pushing it toward benign (1.00). Furthermore, in Fig.~\ref{fig: MAll3} benign traffic is misclassified as malicious traffic because the larger contributing features have values that push it towards malicious (e.g. the features \textit{Inbound} (0), and \textit{Min Packet Length} (0), etc push it toward malicious).

\begin{figure}[h!]%
    \centering
    \subfloat[\label{fig:MFS1}\centering Malicious: Predicted Malicious]{{\includegraphics[width=9cm]{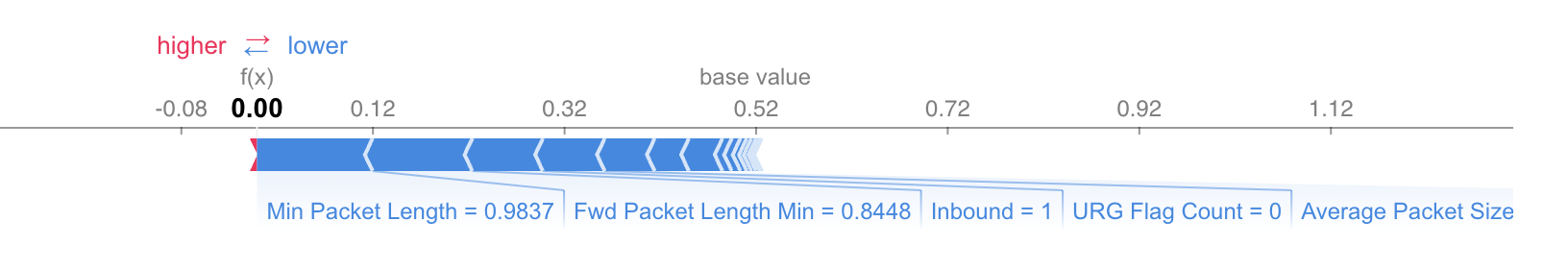} }}%
    % \par
    % \subfloat[\label{fig:MFS2}\centering Benign: Predicted Malicious]{{\includegraphics[width=9cm]{Figures/MFS-10.png} }}%
    \par
    \subfloat[\label{fig:MFS2}\centering Benign: Predicted Benign]{{\includegraphics[width=9cm]{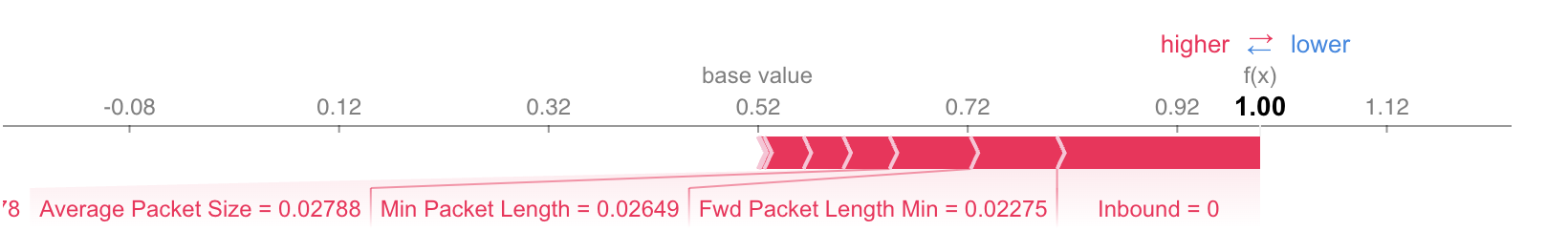} }}%
    \par
    \subfloat[\label{fig:MFS3}\centering Benign: Predicted Malicous]{{\includegraphics[width=9cm]{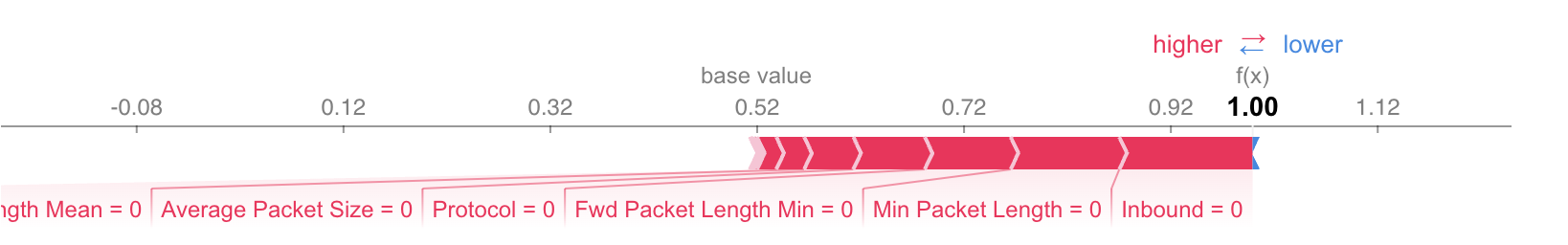} }}%
    \caption{Local Explanation: One-to-all with Feature Selection}%
    \label{fig:Loc-MFS}%
\end{figure}

% \begin{figure*}[t]
%      \centering
%      \begin{subfigure}%[b]{0.55\textwidth}
%      \includegraphics[width=1\linewidth]{Figures/local-one-multiple.png}
%      \caption{One-to-Multiple: Local Explanation of Attacks}
%      \label{fig:loc-one-mul} 
%      \end{subfigure}
     
%      \begin{subfigure}%[b]{0.55\textwidth}
%      \includegraphics[width=1\linewidth]{Figures/local-multiple-attack.png}
%      \caption{One-to-Multiple: Local Explanation of Benign}
%      \label{fig:loc-multiple-attack}
%      \end{subfigure}
% % \caption{Local Explanation}
% \end{figure*}
% \subsection{Comparison to other methods}

\section{Conclusion}\label{sec:conclusion}
In this study, we propose a framework that can efficiently classify legitimate traffic and malicious traffic of DDoS attacks and use Kernel SHAP to provide an interpretation of the decision-making of the prediction results of the MLP classifier. In our proposed framework, we first select the top 20 important features based on three XGB-based feature importance techniques and create a subset of optimized features from the CICDDoS2019 dataset. The MLP classifier can be applied to classify legitimate traffic (benign) and malicious traffic by using the subset data of optimized features. We evaluated two binary classification scenarios: 1) one-to-one classification, which classifies legitimate traffic and a type of malicious traffic; and 2) one-to-all classification, where legitimate traffic and malicious traffic (including multiple types of DNS, LDAP, SNMP, and NetBIOS traffic labeled as malicious) of DDoS attacks are classified. The evaluation results show that the classification model achieves a high performance of over 99\% accuracy in both two scenarios with the top 20 selected important features. To provide interpretability, we use SHAP values to explain the prediction results of the classification, which is important to understand how the MLP classifier model classifies and identifies DDoS traffic. Furthermore, the explanation method aims to: 1) bring the feature contribution in decreasing order from high to low for the prediction results in the global explanation, while the local explanation provides explanations for a specific instance that the feature contribution in both legitimate traffic and malicious traffic separately, 2) present a visual explanation of the most influential features, 3) analyze the feature contribution in both legitimate traffic and malicious traffic separately, and 4) depicts a visualization of the influence of individual features on the prediction results. 

This study provides an efficient model for binary classification and a visual explanation of the decision-making process of the classification model. In future work, we plan to provide an explanation for the multi-class classification\cite{wei2021ae}.

% \section*{Acknowledgments}
% This should be a simple paragraph before the References to thank those individuals and institutions who have supported your work on this article.


% \newpage

 

% \end{IEEEbiography}



\bibliographystyle{IEEEtran}
\bibliography{mybib}

\vfill

\end{document}